\documentclass[12pt,a4paper]{article}
\usepackage[latin9]{inputenc}% This allows you to write é instead of \'e
\usepackage{a4wide}
\usepackage{latexsym}
\usepackage{epsf}
\usepackage{amssymb}
\usepackage{ifpdf}
\ifpdf
\usepackage[pdftex]{graphicx}
\usepackage[pdftex,unicode,implicit]{hyperref}
\hypersetup{%
  pdftitle    = {Supersymmetry, attractors and cosmic censorship.},
  pdfkeywords = {N=2 Supersymmetry, attractors, BPS black holes},
  pdfauthor   = {J. Bellorín, P. Meessen, T. Ortín},
  pdfcreator  = {pdf\LaTeXe\ with package \flqq hyperref\frqq},
  pdfproducer = {pdf\LaTeXe\ with package \flqq hyperref\frqq},
  pdfpagemode = None,  %change this to FullScreen if you want...
  pdffitwindow= true,  % Open the pdf file without the sidebars open.
  unicode     = true,
  plainpages  = false,
  colorlinks  = true,  % links have no different colors..
  citecolor   = black,
  urlcolor    = black,
  linkcolor   = black
}
\newcommand{\hepth}[1]{arXiv:{\tt \href{http://www.arXiv.org/abs/hep-th/#1}{hep-th/#1}}}
\newcommand{\grqc}[1]{arXiv:{\tt \href{http://www.arXiv.org/abs/gr-qc/#1}{gr-qc/#1}}}

\else
  \usepackage[dvips]{graphicx}
  \usepackage[unicode,implicit]{hyperref}
  \newcommand{\hepth}[1]{arXiv:{\tt hep-th/#1}}
  \newcommand{\grqc}[1]{arXiv:{\tt gr-qc/#1}}
  
\fi
   \makeatletter
\@addtoreset{equation}{section}
\makeatother

\begin{document}
%%%%%%
\begin{flushright}
\small
IFT-UAM/CSIC-06-29\\
{\bf hep-th/0606201}\\
%June $21^{\rm st}$, $2006$    % Date version 1
October $19^{\rm th}$, $2006$   % Date revised version.
\normalsize
\end{flushright}
\begin{center}
%title
\vspace{1.5cm}
{\Large {\bf Supersymmetry, attractors\\[.5cm] and\\[.5cm] cosmic censorship}} \vspace{2cm}

%authors
{\sl\large Jorge Bellor\'{\i}n}
\footnote{E-mail: {\tt Jorge.Bellorin@uam.es}},
{\sl\large Patrick Meessen}
\footnote{E-mail: {\tt Patrick.Meessen@cern.ch}},
{\sl\large and Tom{\'a}s Ort\'{\i}n}
\footnote{E-mail: {\tt Tomas.Ortin@cern.ch}}

\vspace{1cm}

{\it Instituto de F\'{\i}sica Te\'orica UAM/CSIC\\
  Facultad de Ciencias C-XVI,  C.U.~Cantoblanco,  E-28049-Madrid, Spain}\\

\vspace{1.5cm}

%%%%%%%%%%%%%%%%%%%%%%%%%%%%%%%%%%%%%%%%%%%%%%%%%%%%%%%%%%%%%%%%%%%%%%

{\bf Abstract}

\end{center}

\begin{quotation}

\small

We show that requiring unbroken supersymmetry \textit{everywhere} in
black-hole-type solutions of $N=2,d=4$ supergravity coupled to vector
supermultiplets ensures in most cases absence of naked singularities.  We
formulate three specific conditions which we argue are equivalent to the
requirement of global supersymmetry. These three conditions can be related to
the absence of sources for NUT charge, angular momentum, scalar hair and
negative energy, although the solutions can still have globally defined
angular momentum and non-trivial scalar fields, as we show in an explicit
example.  Furthermore, only the solutions satisfying these requirements seem
to have a microscopic interpretation in String Theory since only they have
supersymmetric sources.  These conditions exclude, for instance, singular
solutions such as the Kerr-Newman with $M=|q|$, which fails to be everywhere
supersymmetric.

We also present a re-derivation of several results concerning attractors in 
$N=2,d=4$ theories based on the explicit knowledge of the most general
solutions in the timelike class.

\end{quotation}

\newpage

\pagestyle{plain}

%%%%%%%%%%%%%%%%%%%%%%%%%%%%%%%%%%%%%%%%%%%%%%%%%%%%%%%%%%%%%%%%%%%%%%
%%%%%%%%%%%%%%%%%%%%%%%%%%%%%%%%%%%%%%%%%%%%%%%%%%%%%%%%%%%%%%%%%%%%%%
%%%%%%%%%%%%%%%%%%%%%%%%%%%%%%%%%%%%%%%%%%%%%%%%%%%%%%%%%%%%%%%%%%%%%%
%%%%%%%%%%%%%%%%%%%%%%%%%%%%%%%%%%%%%%%%%%%%%%%%%%%%%%%%%%%%%%%%%%%%%%
%%%%%%%%%%%%%%%%%%%%%%%%%%%%%%%%%%%%%%%%%%%%%%%%%%%%%%%%%%%%%%%%%%%%%%

\tableofcontents

%\newpage
%%%%%%%%%%%%%%%%%%%%%%%%%%%%%%%%%%%%%%%%%%%%%%%%%%%%%%%%%%%%%%%%%%%%%%
%%%%%%%%%%%%%%%%%%%%%%%%%%%%%%%%%%%%%%%%%%%%%%%%%%%%%%%%%%%%%%%%%%%%%%
%%%%%%%%%%%%%%%%%%%%%%%%%%%%%%%%%%%%%%%%%%%%%%%%%%%%%%%%%%%%%%%%%%%%%%
%%%%%%%%%%%%%%%%%%%%%%%%%%%%%%%%%%%%%%%%%%%%%%%%%%%%%%%%%%%%%%%%%%%%%%
%%%%%%%%%%%%%%%%%%%%%%%%%%%%%%%%%%%%%%%%%%%%%%%%%%%%%%%%%%%%%%%%%%%%%%

\section{Introduction}

In spite of the impressive progress made during the last few years in the
study of supersymmetric black-hole solutions, there are important questions
that remain unanswered or whose answer is unclear. For instance, we know how
to construct many supersymmetric black-hole-type solutions, but many of them
are singular.  Some of these become regular when string corrections are taken
into account and for all the regular black hole solutions we seem to have a
String Theory model that accounts for its entropy. How are the other singular
solutions to be understood? How can it be that they are supersymmetric and yet
there is no String Theory model for them? Or, if there is, why are they
singular?

The main goal of this paper is to try to answer this question by giving a set
of conditions that supersymmetric black-hole-type solutions must satisfy in
order to be admissible in the context of $N=2,d=4$ supergravity coupled to
vector supermultiplets. Admissible solutions will be regular and will describe
one or several black holes in static equilibrium, even though the system may
have a finite global angular momentum, as is for example the case in the
solution constructed in Ref.~\cite{Elvang:2005sa}. Furthermore, we expect only
admissible solutions to have a miscroscopic String Theory model. We will argue
that the non-admissible solutions are, in general, not truly supersymmetric in
the sense that will be explained later on and the conditions of admissibility
can be seen as conditions for a solution to be everywhere supersymmetric. For
instance: the Kerr-Newman solution with equal charge and mass, which is
singular but nevertheless commonly believed to be supersymmetric, is
non-admissible according to our criteria.  We will show that it fails to be
supersymmetric at the singularity, where the sources might be located.
Equivalently we can say that the Kerr-Newman field with $M=|q|$ is caused by
non-supersymmetric sources. This explains why it is not described by any
supersymmetric String Theory model. We will also show that, generically,
rotating sources are not allowed by supersymmetry and that regular,
supersymmetric solutions with angular momentum are always composite objects
made out of several static black holes in equilibrium. The angular momentum
has its origin in the dipole momenta of the electromagnetic fields
corresponding to the distribution of charged black holes. Something similar
happens for scalar fields: supersymmetric configurations satisfying our
conditions can have non-trivial scalar fields but cannot have sources.

In order to prove these results, we will make use of the explicit knowledge of
the most general solutions of $N=2,d=4$ supergravity coupled to vector
multiplets, which have recently been classified in
Ref.~\cite{Meessen:2006tu}\footnote{In this paper we will not consider the
  coupling to hypermultiplets. The classification of the supersymmetric
  solutions with both vector multiplets and hypermultiplets is considered in
  Ref.~\cite{kn:HMO}.} .  All the asymptotically flat supersymmetric black
hole solutions seem to belong to the timelike class, and, although they
coincide with the solutions found in Ref.~\cite{Behrndt:1997ny}, the general
formalism will allow us to make further progress in their understanding. In
particular, we will use the \textit{Killing Spinor Identities} (KSIs)
\cite{Kallosh:1993wx,Bellorin:2005hy}, which can be understood as
integrability conditions for the Killing spinor equations, in order to study
supersymmetry at the singular points where the sources of these solutions
should be located.

The final ingredient will be the attractor equations of $N=2,d=4$ supergravity
\cite{Ferrara:1995ih,Strominger:1996kf,Ferrara:1996dd,Ferrara:1996um}: these
provide us with information about the sources thought of as being placed at
the attractor points. In fact, we will find interesting relations between KSIs
and attractor equations, the former showing explicitly that

\begin{enumerate}
\item supersymmetry always requires the absence of the kind of scalar hair
  called \textit{primary} in Ref.~\cite{Coleman:1991ku}, and that 
  
\item when the attractor equations are satisfied there are no sources
  whatsoever for scalar hair.\footnote{
    If there is more than one basin of attraction, contrary to what is 
    assumed in this article, this last conclusion might change due to the
    {\em area codes} \cite{Moore:1998pn}.
}
\end{enumerate}

These results can be viewed as an extension of those of
Ref.~\cite{Kallosh:1992ii} in which it was observed that supersymmetry seems
to act as a cosmic censor for static black-hole-type configurations but not
for the stationary ones, such as the Kerr-Newman $M=|q|$ solution.

This paper is organized as follows: in Section~\ref{sec-d4BPS} we review
the timelike class of supersymmetric solutions of $N=2,d=4$ supergravity
coupled to vector multiplets. First, we review how all the solutions in this
class can be constructed from a symplectic vector of real harmonic functions
and then in Section~\ref{sec-KSI} we derive the KSIs that, by assumption of
supersymmetry, all solutions must satisfy. 
In Section~\ref{sec-attractors} a re-derivation of some of the main results
involving $N=2,d=4$ supersymmetric black-hole attractors, 
taking advantage of the
actual and explicit knowledge of all the solutions of this kind, which has
helped us to improve some of the presentations existing in the literature and
prove some new results.

In Section~\ref{sec-n2d4} we study how the KSIs constrain the possible sources
and singularities of black-hole-type solutions and the interplay with the
attractor equations in a general way. The main result of this section will be
the formulation of three conditions that express the existence of
supersymmetry everywhere in the solutions, including, particularly, the
locations of the sources. These conditions should ensure the regularity of the
admissible solutions and we study in very close detail several examples in
Section~\ref{sec-n2d4bhs}.  Section~\ref{sec-conclusion} contains our
conclusions.

%%%%%%%%%%%%%%%%%%%%%%%%%%%%%%%%%%%%%%%%%%%%%%%%%%%%%%%%%%%%%%%%%%%%%%
%%%%%%%%%%%%%%%%%%%%%%%%%%%%%%%%%%%%%%%%%%%%%%%%%%%%%%%%%%%%%%%%%%%%%%
%%%%%%%%%%%%%%%%%%%%%%%%%%%%%%%%%%%%%%%%%%%%%%%%%%%%%%%%%%%%%%%%%%%%%%
%%%%%%%%%%%%%%%%%%%%%%%%%%%%%%%%%%%%%%%%%%%%%%%%%%%%%%%%%%%%%%%%%%%%%%
%%%%%%%%%%%%%%%%%%%%%%%%%%%%%%%%%%%%%%%%%%%%%%%%%%%%%%%%%%%%%%%%%%%%%%
%%%%%%%%%%%%%%%%%%%%%%%%%%%%%%%%%%%%%%%%%%%%%%%%%%%%%%%%%%%%%%%%%%%%%%

\section{Timelike BPS solutions of $N=2,d=4$ SUEGRA}
\label{sec-d4BPS}

It was recently shown in Ref.~\cite{Meessen:2006tu} that all the
supersymmetric solutions in the timelike class of $N=2,d=4$ supergravity
coupled to $n$ vector multiplets\footnote{These solutions were first found in
  slightly different form in Ref.~\cite{Behrndt:1997ny} and the procedure
  followed in Ref~\cite{Meessen:2006tu} shows that they are the only solutions
  in this class.} can be constructed by setting the $2\bar{n}=2(n+1)$
components of a real, symplectic vector
$\mathcal{I}=(\mathcal{I}^{\Lambda},\mathcal{I}_{\Lambda})$ equal to
$2\bar{n}=2(n+1)$ real functions harmonic on 3-dimensional Euclidean
space\footnote{If the functions are not harmonic, the field configurations are
  still supersymmetric, but are {\em not} solutions of the equations of motion.}

\begin{equation}
\label{eq:stabilization}
\mathcal{I} 
\equiv
\left( 
  \begin{array}{c}
  \mathcal{I}^{\Lambda} \\  
  \mathcal{I}_{\Lambda} \\  
  \end{array}
\right)
\, ,
\hspace{1cm}
\partial_{m}\partial_{m} \mathcal{I}^{\Lambda}
=
\partial_{m}\partial_{m} \mathcal{I}_{\Lambda}=0\, ,
\hspace{1cm}
\Lambda=0,1,\cdots,n\, .
\end{equation}

\noindent
This real section $\mathcal{I}$ enters the theory as the imaginary
part of the section $\mathcal{V}/X$, where $\mathcal{V}$ is the
covariantly-holomorphic canonical section defining special geometry:

\begin{equation}
\label{eq:SGDefFund}
\mathcal{V} = 
\left(
\begin{array}{c}
\mathcal{L}^{\Lambda}\\
\mathcal{M}_{\Sigma}\\
\end{array}
\right) \;\; \rightarrow \;\;
\left\{
\begin{array}{lcl}
\langle \mathcal{V}\mid\mathcal{V}^{*}\rangle 
& \equiv & 
\mathcal{L}^{*\, \Lambda}\mathcal{M}_{\Lambda} 
-\mathcal{L}^{\Lambda}\mathcal{M}^{*}_{\Lambda}
= -i\, , \\
& & \\
\mathfrak{D}_{i^{*}}\mathcal{V} & = & (\partial_{i^{*}}-
{\textstyle\frac{1}{2}}\partial_{i^{*}}\mathcal{K})\mathcal{V} =0 \, ,\\
& & \\
\langle\mathfrak{D}_{i}\mathcal{V}\mid\mathcal{V}\rangle & = & 0 \, .
\end{array}
\right. 
\end{equation}

$X$ on the other hand is proportional to the complex, scalar bilinear constructed out
of the Killing spinors: 
supersymmetry and consistency of the solutions imply that it can be expressed
in terms of $\mathcal{I}$, see e.g.~Ref.~\cite{Meessen:2006tu} or Eq.~(\ref{eq:MRI}).

Eqs.~(\ref{eq:stabilization}) are sometimes known as the \textit{generalized
  stabilization equations}, the standard stabilization equations having the
same form but with the harmonic functions
$(\mathcal{I}^{\Lambda},\mathcal{I}_{\Lambda})$ replaced by magnetic and
electric charges, {\em e.g.\/} $(p^{\Lambda},q_{\Lambda})$.

The real part of $\mathcal{V}/X$, denoted by $\mathcal{R}\equiv
(\mathcal{R}^{\Lambda},\mathcal{R}_{\Lambda})$ can, in principle, be written
in terms of the real harmonic functions, which is usually referred to as
``solving the stabilization equations''. In theories with a prepotential, the
homogeneity properties of the prepotential allow us to write

\begin{equation}
\label{eq:MdF}
  \mathcal{M}_{\Lambda}/X =
\frac{\partial \mathcal{F}(\mathcal{L}^{\cdot}/X)}{\partial (\mathcal{L}^{\Lambda}/X)}\, .
\end{equation}

\noindent
Taking the imaginary part of this equation, we have

\begin{equation}
\label{eq:IRHH}
\mathcal{I}_{\Lambda} (\mathcal{R}^{\cdot},\mathcal{I}^{\cdot})
=\mathcal{I}_{\Lambda}\, , 
\end{equation}

\noindent
which implicitly defines
$\mathcal{R}^{\Lambda}(\mathcal{I}^{\cdot},\mathcal{I}_{\cdot})$, although
solving these equations can be extremely hard and in general the 
explicit solution is unknown.

The real part of Eqs.~(\ref{eq:MdF}) and the above solutions give
straightforwardly the functions
$R_{\Lambda}(R^{\cdot}(\mathcal{I}^{\cdot},\mathcal{I}_{\cdot}),
\mathcal{I}^{\cdot})$.

Having the complete symplectic section $\mathcal{V}/X$ entirely given
in terms of the real harmonic functions, one can construct the
fields of the solutions as follows: 

\begin{enumerate}
\item The $n$ complex scalar fields $Z^{i}$ are given by the quotients

  \begin{equation}
\label{eq:scalars}
   Z^{i}= \frac{\mathcal{L}^{i}/X}{\mathcal{L}^{0}/X}=
\frac{\mathcal{R}^{i}+i\mathcal{I}^{i}}{\mathcal{R}^{0}+i\mathcal{I}^{0}}\, .
  \end{equation}

\item The metric has the form

\begin{equation}
\label{eq:metricd4}
ds^{2} = 2|X|^{2}(dt+\omega)^{2} -\frac{1}{2|X|^{2}}dx^{i}dx^{i}\, ,
\hspace{1cm}
i,j=1,2,3\, ,
\end{equation}

\noindent
where 

\begin{equation}
\label{eq:MRI}
\frac{1}{2|X|^{2}}=  \langle\,\mathcal{R}\mid \mathcal{I}\, \rangle\, ,
\end{equation}

\noindent
and $\omega$ is a time-independent 1-form on Euclidean 3-dimensional
space satisfying the equation

\begin{equation}
\label{eq:oidi}
(d\omega)_{mn} =2\epsilon_{mnp}
\langle\,\mathcal{I}\mid \partial_{p}\mathcal{I}\, \rangle\, .  
\end{equation}

\item The symplectic vector of field strengths and their duals
  $F=(F^{\Lambda},\tilde{F}_{\Lambda})$ is given by

\begin{equation}
\label{eq:esas}
F = -{\textstyle\frac{1}{2}} \{d[\mathcal{R} \hat{V}] 
-{}^{\star}[d\mathcal{I}\wedge \hat{V}] \}\, ,
\hspace{1cm}
\hat{V}=2\sqrt{2}|X|^{2}(dt+\omega)\, .
\end{equation}

\end{enumerate}

The Killing spinors of these solutions have the form

\begin{equation}
\epsilon_{I}=X^{1/2}\epsilon_{I\, 0}\, ,
\hspace{1cm}
\partial_{\mu}\epsilon_{I\, 0}=0\, ,
\hspace{1cm}  
\epsilon_{I\, 0} +i\gamma_{0}
  \epsilon_{IJ}\epsilon^{J}{}_{0}=0\, ,  
\end{equation}

\noindent
which implies

\begin{equation}
\label{eq:constraint}
\epsilon_{I} +i\gamma_{0}e^{i\alpha} \epsilon_{IJ}\epsilon^{J}=0\, ,  
\hspace{1cm}
e^{i\alpha}= (X/X^{*})^{1/2}\, .
\end{equation}

Observe that we can write

\begin{equation}
X = \frac{\mathcal{L}^{\Lambda}(Z,Z^{*})}{\mathcal{R}^{\Lambda} 
+i\mathcal{I}^{\Lambda}}\, ,  
\end{equation}

\noindent
for any $\Lambda$.

%%%%%%%%%%%%%%%%%%%%%%%%%%%%%%%%%%%%%%%%%%%%%%%%%%%%%%%%%%%%%%%%%%%%%%
%%%%%%%%%%%%%%%%%%%%%%%%%%%%%%%%%%%%%%%%%%%%%%%%%%%%%%%%%%%%%%%%%%%%%%

\subsection{Killing Spinor Identities}
\label{sec-KSI}

All supersymmetric configurations satisfy the \textit{Killing Spinor
Identities} relating the Einstein equations $\mathcal{E}^{\mu\nu}$, the
Maxwell equations $\mathcal{E}_{\Lambda}{}^{\mu}$, the Bianchi identities
$\mathcal{B}^{\Lambda\, \mu}$ and the scalar equations of motion
$\mathcal{E}^{i}$ \cite{Kallosh:1993wx,Bellorin:2005hy,Meessen:2006tu}

\begin{eqnarray}
\label{eq:ksipsi}
\mathcal{E}_{a}{}^{\mu}\gamma^{a}\epsilon_{I} -4i
\langle\, \mathcal{E}^{\mu} \mid \, \mathcal{V}\, \rangle 
\epsilon_{IJ}\epsilon^{J} & = & 0\, ,\\
& & \nonumber \\
\label{eq:ksilambda} 
\mathcal{E}^{i} \epsilon^{I} +2i
\langle\, \not\!\mathcal{E} \mid \, \mathcal{U}^{*\, i}\, \rangle 
\epsilon^{IJ} \epsilon_{J}& = & 0\, ,
\end{eqnarray}

\noindent
where $\mathcal{E}^{\mu}$ is the symplectic vector $(\mathcal{B}^{\Lambda\,
  \mu}, \mathcal{E}_{\Lambda}{}^{\mu})$. 

In the timelike case, they lead to the following identities in an orthonormal
frame

\begin{eqnarray}
\label{eq:ksi1}
\mathcal{E}^{ab} & = & \eta^{a}{}_{0}  \eta^{b}{}_{0}\mathcal{E}^{00}\, ,\\
& & \nonumber \\
\label{eq:ksi2}
\langle\, \mathcal{V}/X \mid \, \mathcal{E}^{a} \, \rangle  & = & 
{\textstyle\frac{1}{4}}|X|^{-1} \mathcal{E}^{00}\delta^{a}{}_{0}\, ,\\
& & \nonumber \\
\label{eq:ksi3}
\langle\, \mathcal{U}^{*}_{i^{*}}\mid \, \mathcal{E}^{a} \, \rangle  & = & 
{\textstyle\frac{1}{2}} e^{-i\alpha}\mathcal{E}_{i^{*}}\delta^{a}{}_{0}\, .
\end{eqnarray}

\noindent
These equations imply directly

\begin{equation}
\mathcal{E}^{0m}=0\, ,\,\,\,\,\,
\mathcal{E}^{mn}=0\, ,\,\,\,\,\,
\langle\, \mathcal{V}\mid \, \mathcal{E}^{m} \, \rangle =0\, ,\,\,\,\,\,
\langle\, \mathcal{U}^{*}_{i^{*}}\mid \, \mathcal{E}^{m} \, \rangle=0\, .
\end{equation}

\noindent
Further, the r.h.s.~of Eq.~(\ref{eq:ksi1}) is real, and this leads to
two important identities:

\begin{eqnarray}
\label{eq:ksi2-2}
\langle\, \mathcal{I} \mid \, \mathcal{E}^{0} \, \rangle  & = & 0\, ,\\
& & \nonumber \\
\label{eq:ksi2-1}
\mathcal{E}^{00} & = & \pm 4|\langle\, \mathcal{V} \mid \, \mathcal{E}^{0}\,
\rangle|\, .
\end{eqnarray}

%%%%%%%%%%%%%%%%%%%%%%%%%%%%%%%%%%%%%%%%%%%%%%%%%%%%%%%%%%%%%%%%%%%%%%
%%%%%%%%%%%%%%%%%%%%%%%%%%%%%%%%%%%%%%%%%%%%%%%%%%%%%%%%%%%%%%%%%%%%%%

\subsection{Attractor equations}
\label{sec-attractors}

It is well-known that, in general, the scalar fields of the black-hole
solutions of these theories have certain attractor values that depend solely
on the electric and magnetic charges and which are attained at the event
horizons irrespectively of the chosen asymptotic values
\cite{Ferrara:1995ih,Strominger:1996kf}.\footnote{
   If there are multiple attractor regions, it might happen that there is some 
   residual dependency on the asymptotic values. Here we assume there to be
   only one attractor region.
}
The attractor values are those which
extremize a specific function; furthermore, the absolute value squared of the
central charge for the attractor values is essentially the horizon area
\cite{Ferrara:1996dd,Ferrara:1996um}. Here we are going to rederive these
results using our notation and to relate them to the KSIs. We also want to improve
the previous derivations by making explicit use of the knowledge of all the
supersymmetric configurations.

Let us consider single, static, asymptotically flat, spherically symmetric,
black-hole-type solutions of $N=2,d=4$ supergravity coupled to vector
multiplets: they are given by real harmonic functions of the form

\begin{equation}
\label{eq:Iforspherical}
\mathcal{I} = \mathcal{I}_{\infty} +\frac{q}{r}\, ,
\end{equation}

\noindent
which is the general choice compatible with the assumptions. The
metric can be conveniently written in spherical coordinates as

\begin{equation}
ds^{2} = 2|X|^{2}dt^{2} -\frac{1}{2|X|^{2}}[dr^{2}+r^{2}d\Omega^{2}_{(2)}]\, .
\end{equation}

This metric describes black holes if 

\begin{equation}
-g_{rr}=\frac{1}{2|X|^{2}}\stackrel{r\rightarrow \infty}{\longrightarrow} 
1+\frac{2M}{r}\, ,
\end{equation}

\noindent
is always finite for finite $r$, whence $M$, which is the mass, must be positive.
Further, we have to require

\begin{equation}
\frac{1}{2|X|^{2}} \stackrel{r\rightarrow 0}{\longrightarrow} 
\frac{A}{4\pi r^{2}}>0\, ,   
\end{equation}

\noindent
which imposes the existence of
an event horizon with area $A> 0$ at $r=0$ instead of a naked
singularity.

The existence of attractors (fixed points) of the scalar fields follows from
the fact that in supersymmetric configurations, the scalars satisfy first-order
differential equations, as follows immediately from the Killing spinor equations
associated to the gaugino supersymmetry transformation rule:

\begin{equation}
\label{eq:gaugsusyrule}  
\delta_{\epsilon}\lambda^{Ii}  =  
i\not\!\partial Z^{i}\epsilon^{I} 
+\epsilon^{IJ}\not\!G^{i\, +}\epsilon_{J}=0\, .
\end{equation}

\noindent
To derive the needed first-order equations, we first use the time-independence of
the solutions 

\begin{equation}
i\gamma^{m}\partial_{m} Z^{i}\epsilon^{I} 
-4\epsilon^{IJ}G^{i\, +}{}_{0m}\gamma^{m}\gamma^{0}\epsilon_{J}=0\, ,
\end{equation}

\noindent
and then the known constraint Eq.~(\ref{eq:constraint}) as to obtain

\begin{equation}
(\partial_{m} Z^{i} -4e^{i\alpha}G^{i\, +}{}_{0m})
\gamma^{m}\epsilon^{I}=0\, ,\,\,\,\,
\Rightarrow \partial_{m} Z^{i} =4e^{i\alpha}G^{i\, +}{}_{0m}\, .
\end{equation}

\noindent
Going over to curved indices, the equation takes the form

\begin{equation}
\frac{dZ^{i}}{dr} =2\sqrt{2}G^{i\, +}{}_{tr}/X^{*}\, .  
\end{equation}

\noindent
The self-duality of $G^{i\, +}$ allows us to express the $G^{i\, +}{}_{tr}$
component in terms of the $G^{i\, +}{}_{\theta\phi}$:

\begin{equation}
G^{i\, +}{}_{tr}= i({}^{\star}G^{i\, +}){}_{\theta\phi}=
-i\frac{2|X|^{2}}{r^{2}\sin{\theta}} G^{i\, +}{}_{\theta\phi}\, , 
\end{equation}

\noindent
which combined with 

\begin{equation}
G^{i\, +} =\mathcal{T}^{i}{}_{\Lambda}F^{\Lambda\, +}
= 
{\textstyle\frac{i}{2}}
\mathcal{G}^{ij^{*}}
\langle\, \mathfrak{D}_{j^{*}}\mathcal{V}^{*} \mid F\, \rangle
=
{\textstyle\frac{i}{2}}
\mathcal{G}^{ij^{*}}
\mathfrak{D}_{j^{*}}\langle\, \mathcal{V}^{*} \mid F\, \rangle\, ,   
\end{equation}

\noindent
leads to

\begin{equation}
\label{eq:esaecuacion}
\frac{dZ^{i}}{dr} =2\sqrt{2} \frac{X}{r^{2}\sin{\theta}} 
\mathcal{G}^{ij^{*}}
\mathfrak{D}_{j^{*}}\langle\, \mathcal{V}^{*} \mid F_{\theta\phi}\, \rangle \, .  
\end{equation}

Since the form of all the fields in terms of $\mathcal{I}(r)$ is in
principle known, we can try to find a more explicit form for this equation:
using the general form of the vector fields Eq.~(\ref{eq:esas}) and of
$\mathcal{I}(r)$, Eq.~(\ref{eq:Iforspherical}), we find

\begin{equation}
F_{\theta\phi} 
=
{\textstyle \frac{1}{\sqrt{2}}}r^{2}\sin{\theta}\frac{d\mathcal{I}}{dr}
=
-\frac{q}{\sqrt{2}}\sin{\theta}\, .
\end{equation}

\noindent
After substituting this into Eq.~(\ref{eq:esaecuacion}), one ends up with

\begin{equation}
\frac{dZ^{i}}{d\rho} =2 X \mathcal{G}^{ij^{*}}
\mathfrak{D}_{j^{*}}\mathcal{Z}^{*} \, ,
\end{equation}

\noindent
where  $\rho\equiv 1/r$ and where

\begin{equation}
  \mathcal{Z}(Z,q) \equiv \langle\, \mathcal{V} \mid q \, \rangle\, ,  
\end{equation}

\noindent
is the {\em central charge of the theory} \cite{Ceresole:1995jg}. Observe that the
presence of the factor $X$ in the r.h.s.~is crucial for it to have zero global
K\"ahler weight, just as the l.h.s. 
Further observe that the $r$-dependence is only through
the scalars $Z^{i}(r)$!

The r.h.s.~of this system of differential equations depends only on the scalar
fields $Z^{i}$, and, thus, it is an autonomous system of ordinary differential
equations\footnote{The use of the variable $\rho=1/r$ is essential in this
  argument. it is easy to see that the derivatives of the scalar fields of
  typical black-hole solutions w.r.t.~to $r$ do not vanish at $r=0$, while
  their derivatives w.r.t.~$\rho$ do..} that has fixed points $Z^{i}_{\rm
  fix}$ at the values at which the r.h.s.~vanishes

\begin{equation}
\label{eq:attractorequations}
\left. \mathfrak{D}_{i}\mathcal{Z}\right|_{Z^{i}=Z^{i}_{\rm fix}}=0\, .  
\end{equation}

If the solution of this system of equations exists,  it gives the fixed
values of the scalars $Z^{i}_{\rm fix}$ as functions of the electric and
magnetic charges only

\begin{equation}
Z^{i}_{\rm fix}= Z^{i}_{\rm fix}(q)\, ,  
\end{equation}

\noindent
since the asymptotic values (moduli) $Z^{i}_{\infty}$ do not occur in the
above differential equation.  The fixed values are reached by the scalars at
the value $\rho=\infty$, i.e.~$r=0$, which is where the event horizon would
be, as discussed at the beginning of this section and in what follows.

The fixed values may or may not be admissible, i.e.~they may or may not belong to the
definition domain of the complex coordinates $Z^{i}$. If the asymptotic values
$Z^{i}_{\infty}$ are admissible and the fixed values $Z^{i}_{\rm fix}(q)$ are not,
there must be a singularity between $r=\infty$ and $r=0$, which will
induce a curvature singularity. We will require both the asymptotic and 
the fixed values to be admissible. 
These aspects will be discussed in Section~\ref{sec-n2d4}.

Black-hole solutions whose scalars take the asymptotic values
$Z^{i}_{\infty}=Z^{i}_{\rm fix}$ have constant scalar fields, and
are called \textit{doubly extreme black holes}.  These values are the ones
that extremize, not the central charge, but the zero-K\"ahler-weight
combination $e^{\mathcal{K}/2}\mathcal{Z}$:

\begin{equation}
\left. \mathfrak{D}_{i}\mathcal{Z}\right|_{Z^{i}=Z^{i}_{\rm fix}}=
\left.  
e^{-\mathcal{K}/2}
\partial_{i}\left(e^{\mathcal{K}/2}\mathcal{Z}\right)\right|_{Z^{i}=
Z^{i}_{\rm fix}}=0\, .
\end{equation}

%%%%%%%%%%%%%%%%%%%%%%%%%%%%%%%%%%%%%%%%%%%%%%%%%%%%%%%%%%%%%%%%%%%%%%
%%%%%%%%%%%%%%%%%%%%%%%%%%%%%%%%%%%%%%%%%%%%%%%%%%%%%%%%%%%%%%%%%%%%%%
%%%%%%%%%%%%%%%%%%%%%%%%%%%%%%%%%%%%%%%%%%%%%%%%%%%%%%%%%%%%%%%%%%%%%%
%%%%%%%%%%%%%%%%%%%%%%%%%%%%%%%%%%%%%%%%%%%%%%%%%%%%%%%%%%%%%%%%%%%%%%

\subsubsection{Consequences of the existence of attractors}

There are no more scalar fields in the theory, but in the timelike
supersymmetric solutions there is another scalar object\footnote{In previous
  derivations in the literature the absolute value $|X|= e^{U}$ is considered,
  but then the K\"ahler weights and the reality properties of the two sides of
  the equations derived are different.} that satisfies a first-order
differential equation: $X$. From the Killing spinor equation associated to the
gravitino supersymmetry transformation rule it is possible to derive
\cite{Meessen:2006tu}

\begin{equation}
\mathfrak{D}_{\mu}X = -i T^{+}{}_{\mu\nu}V^{\nu}\, ,
\end{equation}

\noindent
where $V^{\mu}$ is the timelike Killing vector constructed from the Killing spinor. 
The graviphoton field strength can be written in the form 

\begin{equation}
T^{+} = \langle\, \mathcal{V} \mid F \, \rangle\, ,
\end{equation}

\noindent
and, together with

\begin{equation}
V^{\nu}F_{\nu\mu}=2 \nabla_{\mu}(|X|^{2}\mathcal{R})\, ,
\end{equation}

\noindent
the equation for $X$ becomes

\begin{equation}
\mathfrak{D}_{\mu}X =2 i  \langle\, \mathcal{V} \mid 
\nabla_{\mu}(|X|^{2}\mathcal{R}) \, \rangle\, .
\end{equation}

\noindent
Dividing both sides by $X$ and expanding the r.h.s.~using
$\mathcal{V}/X=\mathcal{R}+i\mathcal{I}$ we get 

\begin{equation}
\frac{\mathfrak{D}_{\mu}X}{X} =2 i |X|^{2} \langle\, \mathcal{R} \mid 
\nabla_{\mu} \mathcal{R} \, \rangle 
-2\nabla_{\mu}|X|^{2} \langle\, \mathcal{I} \mid \mathcal{R} \, \rangle 
-2|X|^{2} \langle\, \mathcal{I} \mid \nabla_{\mu} \mathcal{R} \, \rangle\, . 
\end{equation}

Now, from Eq.~(\ref{eq:MRI}) 

\begin{equation}
-2 \langle\, \mathcal{I} \mid \nabla_{\mu} \mathcal{R} \, \rangle =  
 \nabla_{\mu}|X|^{-2} -2 
\langle\, \mathcal{R} \mid \nabla_{\mu} \mathcal{I} \, \rangle\, ,
\end{equation}

\noindent
and we get

\begin{equation}
\frac{\mathfrak{D}_{\mu}X}{X} =2 i |X|^{2} \langle\, \mathcal{R} \mid 
\nabla_{\mu} \mathcal{R} \, \rangle 
-2|X|^{2} \langle\, \mathcal{R} \mid \nabla_{\mu} \mathcal{I} \, \rangle\, . 
\end{equation}

\noindent
Finally, using 

\begin{equation}
\label{eq:RdR=IdI}
 \langle\, \mathcal{R} \mid 
\nabla_{\mu}\mathcal{R} \, \rangle =
 \langle\, \mathcal{I} \mid 
\nabla_{\mu}\mathcal{I} \, \rangle\, ,  
\end{equation}

\noindent
which is proved in Appendix~\ref{sec-proofs}, we arrive at\footnote{Observe
  that the compatibility between Eq.~(\ref{eq:MRI}) and the following equations
  requires the identity
\begin{equation}
\label{eq:dRI=RdI}
 \langle\, \nabla_{\mu}\mathcal{R} \mid 
\mathcal{I} \, \rangle =
 \langle\, \mathcal{R} \mid 
\nabla_{\mu}\mathcal{I} \, \rangle\, ,  
\end{equation}
\noindent
to hold. For theories admitting a prepotential, this is done in
Appendix~\ref{sec-proofs}.}

\begin{equation}
\label{eq:dxZ}
\mathfrak{D}_{\mu}X^{-1} =2  \langle\, \mathcal{V}^{*} \mid 
\nabla_{\mu} \mathcal{I} \, \rangle\, . 
\end{equation}

\noindent
This equation is valid for all supersymmetric configurations in the timelike
class. For those considered in this section we arrive at the equation we were
looking for:

\begin{equation}
\mathfrak{D}_{\rho}X^{-1} =2  \mathcal{Z}^{*}\, .
\end{equation}

The real and imaginary parts of this equation are

\begin{eqnarray}
\label{eq:dgrr}
\frac{d(-g_{rr})}{d\rho} & = & 2  \Re{\rm e}(\mathcal{Z}^{*}/X^{*})
=2 \langle\, \mathcal{R} \mid q \, \rangle\, ,\\
& & \nonumber \\ 
\frac{d\alpha}{d\rho} +\mathcal{Q}_{\rho} & = & |X|^{2}
-2\Im{\rm m}(\mathcal{Z}^{*}/X^{*})
=2 \langle\, \mathcal{I} \mid q \, \rangle
=2 \langle\, \mathcal{I}_{\infty} \mid q \, \rangle\, .
\end{eqnarray}

\noindent
For the spherically symmetric solutions under consideration $\omega$
vanishes and this requires the phase of $X$ to be covariantly constant, i.e.

\begin{equation}
\label{eq:noNUT}
\langle\, \mathcal{I} \mid q \, \rangle=
\langle\, \mathcal{I}_{\infty} \mid q \, \rangle=0\, .  
\end{equation}

\noindent
We will later show that this is equivalent to the requirement that the
NUT charge vanishes. 
Since there is only dependence on $\rho$, the phase of $X$ can
simply be gauged away by means of a K\" ahler transformations. The phase of
$\mathcal{Z}$ is then also constant, whence $\mathcal{Z}/X$ is real, which 
can be used to write

\begin{equation}
\label{eq:dX-1}
\frac{d |X|^{-1}}{d\rho} =\pm 2  |\mathcal{Z}|\, .
\end{equation}

\noindent
The $\pm$ sign is the sign of $\langle\, \mathcal{R} \mid q \, \rangle$ and we can
argue that it has to be positive if the mass is going to be positive: if we
take Eq.~(\ref{eq:dgrr}) at $\rho=0$ ($r=\infty$), we find that the mass of
the solution is given by the linear combination of charges and moduli

\begin{equation}
\label{eq:linearBPSmass}
M=  \langle\, \mathcal{R}_{\infty} \mid q \, \rangle\, .  
\end{equation}

\noindent
Observe that there is no {\em a priori} guarantee that $M>0$: 
this is a condition that has
to be imposed independently as to avoid singularities. We will do so and will
only consider the positive sign above;
Eq. (\ref{eq:dX-1}) is then the expression found in the literature.
 
If we take another derivative of Eq.~(\ref{eq:dgrr}) and use
Eq.~(\ref{eq:dX-1}), we find

\begin{equation}
\frac{d^{2}(-g_{rr})}{d\rho^{2}} =
2\frac{d |X^{-1}|}{d\rho}|\mathcal{Z}| +2|X|^{-1}\frac{d|\mathcal{Z}|}{d\rho}=
4|\mathcal{Z}|^{2} +2|X|^{-1}\left(\frac{dZ^{i}}{d\rho}\partial_{i}|\mathcal{Z}|
+{\rm c.c.}\right)\, .
\end{equation}

\noindent
Now, at $\rho=\rho_{\rm fix}=0$ we have $Z^{i}=Z_{\rm fix}$ and
$dZ^{i}/d\rho=0$, and the above equation takes on the form

\begin{equation}
\label{eq:AZfix}
\frac{A}{2\pi} = 4|\mathcal{Z}_{\rm fix}|^{2}\, .  
\end{equation}

\noindent
Again, there is no {\em a priori} guarantee that $|\mathcal{Z}_{\rm fix}|\neq 0$, 
which therefore
is another condition that has to be imposed independently as to avoid
singularities.  Actually, even though in this expression $A$ is basically an
absolute value, the positivity of $A$ is only guaranteed if the scalar fields
take admissible values, the mass is positive etc.

These identities allow us to find two interesting expressions for
$|\mathcal{Z}_{\rm fix}|$. Expanding the two sides of Eq.~(\ref{eq:dgrr}) as a
power series in $\rho$ we find 

\begin{equation}
\frac{A}{2\pi} = 
2 \langle\, \left.\frac{d\mathcal{R}}{d\rho}\right|_{\rho=0} 
\mid q \, \rangle\, .  
\end{equation}

\noindent
Using the expressions in Appendix~\ref{sec-proofs} we get 
\cite{Ceresole:1995ca,Ferrara:1996dd}

\begin{equation}
|\mathcal{Z}_{\rm fix}|^{2}
={\textstyle\frac{1}{2}}
\langle\, \left.\frac{d\mathcal{R}}{d\rho}\right|_{\rho=0} 
\mid q \, \rangle  
=-{\textstyle\frac{1}{2}}
q^{T}\mathcal{M}(\mathcal{F}_{\rm fix})q\, ,
\end{equation}

\noindent
where

\begin{equation}
\mathcal{M}(\mathcal{F}) \equiv
\left(
\begin{array}{cc}
\Im{\rm m}\mathcal{F} 
+\Re{\rm e}\mathcal{F}\Im{\rm m}\mathcal{F}^{-1}\Re{\rm e}\mathcal{F} & 
-\Im{\rm m}\mathcal{F}^{-1}\Re{\rm e}\mathcal{F} \\
& \\
-\Re{\rm e}\mathcal{F}\Im{\rm m}\mathcal{F}^{-1} & \Im{\rm m}\mathcal{F}^{-1}  \\
\end{array}
\right)\, .  
\end{equation}

\noindent
A direct computation of $|\mathcal{Z}_{\rm fix}|^{2}$ gives

\begin{equation}
|\mathcal{Z}_{\rm fix}|^{2} = 
|\langle\, \mathcal{V}_{\rm fix} \mid q \, \rangle|^{2}
=-\langle\, q \mid \mathcal{V}_{\rm fix}\, \rangle
\langle\, \mathcal{V}^{*}_{\rm fix}\mid q\, \rangle\, .
\end{equation}

\noindent
The matrix of this bilinear is

\begin{equation}
\mid \mathcal{V}_{\rm fix}\, \rangle
\langle\, \mathcal{V}^{*}_{\rm fix}\mid =
-\left( 
  \begin{array}{cc}
    \mathcal{M}_{\Lambda}\mathcal{M}^{*}_{\Sigma} &  
-\mathcal{M}_{\Lambda}\mathcal{L}^{*\, \Sigma} \\
& \\
    -\mathcal{L}^{\Lambda}\mathcal{M}^{*}_{\Sigma} &  
\mathcal{L}^{\Lambda}\mathcal{L}^{*\, \Sigma} \\
  \end{array}
\right)_{\rm fix}\, .
\end{equation}

\noindent
We can use the relation

\begin{equation}
\mathcal{L}^{*\, \Lambda}\mathcal{L}^{\Sigma}
=
-\textstyle{1\over 2}\Im{\rm m}(\mathcal{N})^{-1|\Lambda\Sigma}
-f^{\Lambda}{}_{i}\mathcal{G}^{ii^{*}}
f^{*\, \Sigma}{}_{i^{*}}\, ,  
\end{equation}

\noindent
taking into account that at the fixed point the second term in the r.h.s.~will
not contribute, and that only its symmetric part will contribute, to
get \cite{Ceresole:1995ca,Ferrara:1996dd}

\begin{equation}
|\mathcal{Z}_{\rm fix}|^{2}
=-{\textstyle\frac{1}{2}}
q^{T}\mathcal{M}(\mathcal{N}_{\rm fix})q\, .
\end{equation}

So far we have checked that the coefficient of the $\rho^{2}$ term of
$-g_{rr}$ is given by the value of the central charge at the fixed point but,
if there are terms of higher order in $\rho$ in $-g_{rr}$ there will not be a
regular horizon. We can, however, see that taking another derivative of
$-g_{rr}$ w.r.t.~$\rho$ at $\rho=0$ will give zero if the attractor equations
(\ref{eq:attractorequations}) are satisfied and the same will happen for
higher derivatives.

Summarizing we can say that the attractor equations 
(plus the positivity of the mass, which is not
guaranteed) seem to be sufficient conditions to have regular, static,
spherically symmetric black holes.

Finally, observe that Eq.~(\ref{eq:linearBPSmass}) plus the identification,
which will be established later on, between the NUT charge and the linear
expression of the charges

\begin{equation}
\label{eq:linearBPSNUT}
N=  \langle\, \mathcal{I}_{\infty} \mid q \, \rangle\, ,    
\end{equation}

\noindent
lead to a complex BPS relation

\begin{equation}
\label{eq:complexlinearBPS}
M+iN= \langle\, (\mathcal{V}/X)_{\infty} \mid q \, \rangle\, .    
\end{equation}

We will argue that supersymmetry requires $N$ to vanish, whence
the above relation reads

\begin{equation}
M= \pm\sqrt{2} |\mathcal{Z}_{\infty}|\, ,
\end{equation}

\noindent
which is the standard BPS relation between mass and central charge. 
Of course, only the positive sign will be admissible.

%%%%%%%%%%%%%%%%%%%%%%%%%%%%%%%%%%%%%%%%%%%%%%%%%%%%%%%%%%%%%%%%%%%%%%
%%%%%%%%%%%%%%%%%%%%%%%%%%%%%%%%%%%%%%%%%%%%%%%%%%%%%%%%%%%%%%%%%%%%%%
%%%%%%%%%%%%%%%%%%%%%%%%%%%%%%%%%%%%%%%%%%%%%%%%%%%%%%%%%%%%%%%%%%%%%%
%%%%%%%%%%%%%%%%%%%%%%%%%%%%%%%%%%%%%%%%%%%%%%%%%%%%%%%%%%%%%%%%%%%%%%
%%%%%%%%%%%%%%%%%%%%%%%%%%%%%%%%%%%%%%%%%%%%%%%%%%%%%%%%%%%%%%%%%%%%%%

\section{Relations between the $N=2,d=4$ KSIs, attractors and sources}
\label{sec-n2d4}

The equations of motion\footnote{By \textit{equation of motion}
  $\mathcal{E}(\phi)$ of a given field $\phi$ we will mean here the l.h.s.~of
  the equation of motion $\delta S/\delta\phi=\mathcal{E}(\phi)=0$. This
  slight abuse of language should lead to no confusions.}  for supersymmetric
configurations of supergravity theories satisfy certain relations known as
\textit{Killing spinor identities} (\textit{KSI}s), which can also be derived
from the integrability conditions of the Killing spinor equations
\cite{Kallosh:1993wx,Bellorin:2005hy}. We have unbroken supersymmetry
wherever the Killing spinors exist, and these exist, locally, wherever the KSIs
are satisfied. Thus, if we are to have unbroken supersymmetry everywhere we
must demand the KSIs to be satisfied everywhere. In this section we are
going to study the consequences of demanding the black-hole solutions of
$N=2,d=4$ supergravity to be everywhere supersymmetric.

The KSIs of $N=2,d=4$ supergravity are given in Eqs.~(\ref{eq:ksipsi}) and
(\ref{eq:ksilambda}) and they lead
to Eqs.~(\ref{eq:ksi1})-(\ref{eq:ksi2-1}) for configurations in the timelike class.
Since we are going to consider configurations that solve the equations of
motion, it may seem that the KSIs are automatically satisfied. However, most
solutions have singularities at which the equations of motion are not
satisfied, i.e.~one has $\mathcal{E}(\phi)=\mathcal{J}(\phi)$.
The r.h.s.~of the
equations of motion at the singularities can be associated to sources for the
corresponding fields and the KSIs are then understood as relations between
the possible sources of supersymmetric solutions: the KSIs put constraints
on possible sources of supersymmetric solutions.

Let us consider from this point of view the KSIs
Eqs.~(\ref{eq:ksi1})-(\ref{eq:ksi2-1}): the first of them,
Eq.~(\ref{eq:ksi1}), tells us that the components $\mathcal{E}^{0m}$ and
$\mathcal{E}^{mn}$ of the Einstein equations must vanish automatically for
supersymmetric configurations and they must do so everywhere if the solutions
are everywhere supersymmetric.  This means that the sources $\mathcal{J}^{0m}$
and $\mathcal{J}^{mn}$ of the Einstein equation must vanish identically
everywhere

\begin{equation}
\label{eq:J0mJmn}
\mathcal{J}^{0m}=\mathcal{J}^{mn}=0\, .  
\end{equation}

\noindent
Hence, singular (delta-like) sources are not allowed, and in particular this means that
no localized sources of angular momentum are allowed.

Any singular contributions to $\mathcal{J}^{0m}$ and
$\mathcal{J}^{mn}$ must originate in the $R^{0m}$ components of the
Ricci tensor; more precisely, they come from the term
$\partial_{\underline{m}}(d\omega)_{\underline{m}\underline{n}}$,
where $\omega$ is the 1-form that appears off-diagonally in the metric
of the timelike supersymmetric solutions of $N=2,d=4$ supergravity
Eq.~(\ref{eq:metricd4}).  Therefore, using Eq.~(\ref{eq:oidi})
and defining the complex 3-dimensional vector
$\vec{\mathcal{W}}$

\begin{equation}
\vec{\mathcal{W}}=
(\mathcal{W}_{\underline{m}})\equiv 
(\langle\, \mathcal{V}/X \mid \, 
\partial_{\underline{m}}\mathcal{I} \, \rangle)\, ,
\hspace{1cm}
\Im{\rm m}\, (\mathcal{W}_{\underline{m}})=
{\textstyle\frac{1}{4}}
\epsilon_{mnp}
(d\omega)_{\underline{n}\underline{p}}
=\langle\, \mathcal{I} \mid \, 
\partial_{\underline{m}}\mathcal{I} \, \rangle\, ,
\end{equation}

\noindent
we can translate the above KSIs, Eqs.~(\ref{eq:J0mJmn}), to the
condition

\begin{equation}
\Im{\rm m}\, (\vec{\nabla}\times \vec{\mathcal{W}})=0\, ,
\end{equation}

\noindent
which has to be imposed everywhere. Actually, only the singular parts of this
equation have to be taken into account since, dealing with solutions, the
finite parts must be canceled in the equations of motion by other finite
contributions. Therefore, from now on we will ignore all finite contributions
to this equation.

Let us consider the real and imaginary parts of Eq.~(\ref{eq:ksi2}), 
namely Eq.~(\ref{eq:ksi2-1}) and (\ref{eq:ksi2-2}). The real part gives us two important
pieces of information: first, it tells us that the component
$\mathcal{J}^{00}$ of the source of the Einstein equation is related to
component $\mathcal{J}^{0}$ of the source of the combined Maxwell and Bianchi
equations $\mathcal{E}^{a}$.  If the electromagnetic fields have only one
static point-like source at $r=0$, $\mathcal{E}^{t}\sim
\frac{1}{\sqrt{2}}q\delta^{(3)}(\vec{x})/ \sqrt{|g|}$, then using the fact that
$\mathcal{Z}/X$ is real (see Eq.~(\ref{eq:noNUT}) and the previous discussion)

\begin{equation}
 \mathcal{E}^{0t}= \pm 2\sqrt{2} \left.|\mathcal{Z}|\right|_{r=0}
\delta^{(3)}(\vec{x})/ \sqrt{|g|}\, ,  
\end{equation}

\noindent
which shows that, if the attractor equations are satisfied, the source for the
Einstein equations is just $\pm |\mathcal{Z}_{\rm fix}(q)|$.  The sign is
related to the positivity of $\langle\, \mathcal{R} \mid q \, \rangle$, which
is, as was discussed before, 
associated to the positivity of the mass etc. This is the only
value admissible by supersymmetry, since we can understand this source as a
source of energy. However, if the scalars take non-admissible values we will
find the wrong sign or a zero at $r=0$ and supersymmetry will be broken at the
source: we will have to require that the attractor equations are solved
by admissible values of the scalars.

The second piece of information we can obtain from the real part concerns the
spacelike components of the electromagnetic sources. Combined with the
spacelike components of the imaginary part, Eq.~(\ref{eq:ksi2-2}), we get the
condition
 
\begin{equation}
\label{eq:VJm}
\langle\, \mathcal{V}/X \mid \, \mathcal{J}^{\underline{m}} \, \rangle =0\, .
\end{equation}

Let us now consider the time component of the imaginary part of the KSI
Eq.~(\ref{eq:ksi2}), Eq.~(\ref{eq:ksi2-2}):

\begin{equation}
\label{eq:IJt}
\langle\, \mathcal{I} \mid \, \mathcal{J}^{t} \, \rangle= 0\, .
\end{equation}

To find the physical meaning of this condition we use the explicit form of the
symplectic vector of vector field strengths $F$ for timelike BPS solutions
Eq.~(\ref{eq:esas}):

\begin{equation}
\label{eq:jotamu}
\mathcal{J}^{\mu}=
\mathcal{E}^{\mu}=
 -({}^{\star}dF)^{\mu}= |X|^{2}\, 
(\partial_{\underline{m}}\partial_{\underline{m}}\mathcal{I})\,  V^{\mu}=
\frac{\delta^{\mu}{}_{t}}{\sqrt{2}}
\frac{\partial_{\underline{m}}\partial_{\underline{m}}\mathcal{I}}{\sqrt{|g|}}\, . 
\end{equation}

\noindent
This result tells us that the KSIs Eq.~(\ref{eq:VJm}) are always satisfied and
that the KSI Eq.~(\ref{eq:IJt}) is equivalent to the condition

\begin{equation}
\label{eq:iddi}
\langle\, \mathcal{I} \mid \, 
\partial_{\underline{m}}\partial_{\underline{m}}\mathcal{I} \, \rangle
=
\Im{\rm m}\, (\partial_{\underline{m}}\mathcal{W}_{\underline{m}})
= 0\, ,
\end{equation}

\noindent
which is nothing but the integrability condition for the 
equation determining $\omega$, which now has to be satisfied everywhere as a consequence of
demanding unbroken supersymmetry \textit{everywhere}. For the point-like
sources considered above, these equations take the form

\begin{equation}
\sum_{A}\langle\, \mathcal{I} \mid \, q_{A} \, \rangle 
\delta^{(3)}(\vec{x}-\vec{x}^{A})/ \sqrt{|g|}=0\, .   
\end{equation}

The consequences of imposing this condition were first studied by
Denef and Bates in Refs.~\cite{Denef:2000nb,Bates:2003vx} in the context of
general $N=2,d=4$ supergravity, but was studied earlier by Hartle and Hawking
in Ref.~\cite{Hartle:1972ya} in the context of Israel-Wilson-Perj\'es (IWP)
solutions of the Einstein-Maxwell theory.  As shown by Tod in
Ref.~\cite{Tod:1983pm} these are precisely the timelike solutions of pure
$N=2,d=4$ supergravity and a special case of the general problem that
we are going to study. 
Hartle and Hawking were motivated, not by supersymmetry, but rather by the
prospect of finding regular solutions describing more than one black hole.  They
were, in particular, worried about possible string singularities related to
NUT charges. These singularities can be eliminated by compactifying the time
coordinate with certain period \cite{kn:M}, but at the price of losing
asymptotic flatness. Let us consider a possible string singularity parametrized
by $z$ and choose polar coordinates $\rho,\phi$ around it.  If one considers the
integral of the 1-form $\omega$ that appears in the metric along a loop of
radius $R$ enclosing the possible string singularity at two different points
$z_{1}$ and $z_{2}$, denoted by $I(R,z_{1,2})$, one can use Stokes' theorem
to derive

\begin{equation}
I(R,z_{1})-I(R,z_{2}) = \int_{\Sigma^{2}} d\omega = 2\int_{\Sigma^{2}}
\star_{3}\Im{\rm m}\mathcal{W}\, ,
\end{equation}

\noindent
where $\Sigma^{2}$ is a surfaces whose boundaries are the loops of radius $R$
at $z_{1,2}$. In the zero radius limit $\Sigma^{2}$ is a closed surface that
crosses the possible string singularity at $z_{1}$ and $z_{2}$ and we have

\begin{equation}
%\lim_{R\rightarrow 0}[I(R,z_{1})-I(R,z_{2})]=
2\pi \lim_{R\rightarrow 0} R [\omega_{\phi}(R,z_{1})-\omega_{\phi}(R,z_{2})]
= 2\int_{\Sigma^{2}}
\star_{3}\Im{\rm m}\mathcal{W}= \int_{\Sigma^{3}}
d\star_{3}\Im{\rm m}\mathcal{W}= 2\int_{\Sigma^{3}} d^{3}x 
\Im{\rm m}\, (\partial_{\underline{m}}\mathcal{W}_{\underline{m}})\, ,
\end{equation}

\noindent
where $\partial\Sigma^{3}=\Sigma^{2}$. Thus, $\Im{\rm m}\,
(\partial_{\underline{m}}\mathcal{W}_{\underline{m}})\neq 0$ implies that
$\omega_{\phi}$ is singular on the string somewhere between $z_{1}$ and
$z_{2}$. These singularities are related to the presence of NUT sources, since
we can define the NUT charge contained in $\Sigma^{3}$ as the integral of
$d\omega$ over $\Sigma^{2}=\partial\Sigma^{3}$:

\begin{equation}
-8\pi N_{\Sigma}= \int_{\Sigma^{2}}d\omega = \int_{\Sigma^{3}}d^{2}\omega = 
2\int_{\Sigma^{3}} d^{3}x 
\Im{\rm m}\, (\partial_{\underline{m}}\mathcal{W}_{\underline{m}})\, .
\end{equation}

%% Observe that the first integral is not zero because $\omega$, just
%% like the electromagnetic potential of a Dirac monopole, is either not
%% globally defined on $S^{2}_{\infty}$ or has singularities. 

Thus, the condition $\Im{\rm m}\,
(\partial_{\underline{m}}\mathcal{W}_{\underline{m}})=0$, required by
supersymmetry, is equivalent to the absence of sources of NUT charge.

Hartle and Hawking argued that the only solutions in the IWP class with no NUT
charge (and no singularities) were the Majumdar-Papapetrou solutions
\cite{Majumdar:1947eu,kn:Pa} which are regular and static.  We will review
their arguments in Section~\ref{sec-twobhsouren2sugra} and show that
there are indeed non-trivial solutions that satisfy the KSIs and have no NUT
charges, apart from the Majumdar-Papapetrou ones; they all have negative
total mass, which causes other naked singularities to appear.

Thus, if we include positivity of all masses among the requirements necessary
to have supersymmetry, the only supersymmetric black-holes-type solutions of
pure $N=2,d=4$ supergravity will indeed be the Majumdar-Papapetrou solutions.
We will have to consider more general $N=2,d=4$ theories in order to be able
to have stationary solutions such as the one found in
Ref.~\cite{Elvang:2005sa}, that satisfy the KSIs and have positive mass. This
will be done in Section~\ref{sec-generaln2d4}.

Next, let us consider the KSI Eq.~(\ref{eq:ksi3}) which relates the sources of
the scalar fields with those of the vector fields. If we consider only
point-like sources and call $\Sigma_{A}$ the scalar charge at $\vec{x}_{A}$,
this equation implies, at each sources

\begin{equation}
\Sigma_{A}=2e^{-i\alpha}\left.\mathfrak{D}_{i}\mathcal{Z}\right|_{\vec{x}_{A}}\, .
\end{equation}

\noindent
As mentioned before, the scalar sources are completely determined by the
electric and magnetic charges and the asymptotic values of the scalar fields.
This is known as \textit{secondary scalar hair} \cite{Coleman:1991ku}. Primary
scalar hair correspond to completely free parameters as in the Einstein-scalar
solutions of Ref.~\cite{kn:JNW} or in the solutions of
Ref.~\cite{Agnese:1994zx} which may be embedded in $N=4,d=4$ supergravity.
Neither of these solutions is supersymmetric (nor regular) and the above KSI
explains just why. 

But there is more to the above KSI: it shows that the existence of
attractors at the sources implies total absence of scalar sources, either of
primary or secondary type. Since this seems to be necessary in order to have
regular event horizons, this KSI implies that there will not be supersymmetric
black holes with scalar hair in these theories. Unfortunately, it seems
possible to have singular supersymmetric solutions with primary scalar hair.

We can summarize the results obtained in this section as follows: we have
identified a series of requirements necessary to avoid singularities in
supersymmetric black-hole-type solutions of $N=2,d=4$ supergravity coupled to
vector multiplets, which can be associated to having unbroken supersymmetry
everywhere (including the sources).

\begin{itemize}
\item[\textbf{I}] The conditions

\begin{eqnarray}
\label{eq:dW}
\Im{\rm m}\,  (\vec{\nabla}\times \vec{\mathcal{W}})
& =  & 0\, ,\\
& & \nonumber \\
\label{eq:d*W} 
\Im{\rm m}\,  (\vec{\nabla}\cdot \vec{\mathcal{W}})
& = & 0\, , 
\end{eqnarray}

\noindent
have to be satisfied everywhere in order to have supersymmetry everywhere.
They ensure the absence of string singularities associated to source of NUT
charge and other singularities associated to sources of angular momentum We
stress that, when dealing with solutions, all finite contributions to the
first equation should be ignored and the second equation can only have
singular terms in the l.h.s.

\item[\textbf{II}] The mass has to be positive. Actually, the masses of each
  of the sources of the solutions should be positive. They cannot be
  rigorously defined in general (for multi-black-hole solutions), but they can
  be identified with certain confidence in the supersymmetric configurations
  at hands \cite{kn:BrilLin}.
  
\item[\textbf{III}] The attractor equations (\ref{eq:attractorequations}) must
  be satisfied at each of the sources for admissible values of the scalars and
  the value of the central charge at each of them must be finite. As we have
  seen, the first condition is equivalent to the total absence of scalar sources. 

\end{itemize}

The last two conditions are associated to the finiteness and positivity of
$-g_{rr}$ outside the sources.  Since $-g_{rr}\sim e^{-\mathcal{K}}$, it
would be finite and positive as long as the scalar fields take admissible
values within their domain of definition. All the zeroes of $-g_{rr}$ can be
related to singularities of the scalar fields. Imposing that the scalar fields
take admissible values everywhere is too strong a condition, since it is
almost equivalent to directly impose absence of singularities in the metric.

The conditions that we have imposed are, however, heuristically equivalent:
for a single black-hole solution the conditions of asymptotic flatness and
positivity of the masses ensure positivity of $-g_{rr}$ in the limit
$r\rightarrow \infty$. The third condition ensures positivity in the
$r\rightarrow 0$ limit and, furthermore, ensures that there will be a horizon
of finite area. Since there are no reasons to expect singularities at finite
values of $r$, the positivity and finiteness should hold for all finite values
of $r$. The same should happen in multi-black-hole solutions.

%%%%%%%%%%%%%%%%%%%%%%%%%%%%%%%%%%%%%%%%%%%%%%%%%%%%%%%%%%%%%%%%%%%%%%
%%%%%%%%%%%%%%%%%%%%%%%%%%%%%%%%%%%%%%%%%%%%%%%%%%%%%%%%%%%%%%%%%%%%%%
%%%%%%%%%%%%%%%%%%%%%%%%%%%%%%%%%%%%%%%%%%%%%%%%%%%%%%%%%%%%%%%%%%%%%%
%%%%%%%%%%%%%%%%%%%%%%%%%%%%%%%%%%%%%%%%%%%%%%%%%%%%%%%%%%%%%%%%%%%%%%
%%%%%%%%%%%%%%%%%%%%%%%%%%%%%%%%%%%%%%%%%%%%%%%%%%%%%%%%%%%%%%%%%%%%%%

\section{$N=2,d=4$ attractors, KSIs and BPS black-hole sources}
\label{sec-n2d4bhs}

Now we want to apply the results of the previous sections to several examples
of black-hole-type solutions of $N=2,d=4$ supergravity theories, demanding the
three conditions formulated in the introduction and checking the regularity of
those solutions that satisfy them. We are going to start with the simplest
theory.

%%%%%%%%%%%%%%%%%%%%%%%%%%%%%%%%%%%%%%%%%%%%%%%%%%%%%%%%%%%%%%%%%%%%%%
%%%%%%%%%%%%%%%%%%%%%%%%%%%%%%%%%%%%%%%%%%%%%%%%%%%%%%%%%%%%%%%%%%%%%%

\subsection{Pure $N=2,d=4$ supergravity}
\label{sec-puren2d4}

This theory has $\bar{n}=1$, no scalar fields, and it is given by the
prepotential

\begin{equation}
\label{eq:prepotentialpure}
\mathcal{F}=-{\textstyle\frac{i}{2}} (\mathcal{X}^{0})^{2}\, ,\,\,\,
\Rightarrow \mathcal{F}_{0}=-i\mathcal{X}^{0}\, . 
\end{equation}

\noindent
This implies that the components of the symplectic section $\mathcal{V}$ are
constant

\begin{equation}
\mathcal{L}^{0}=i\mathcal{M}_{0}=e^{i\gamma}/\sqrt{2}\, ,  
\end{equation}

\noindent
and $X$ is not related to any K\"ahler potential, but 

\begin{equation}
X = \frac{e^{i\gamma}}{\sqrt{2}}(\mathcal{L}^{0}/X)^{-1}
=  \frac{e^{i\gamma}}{\sqrt{2}(\mathcal{R}^{0}+i\mathcal{I}^{0})}\, .    
\end{equation}

\noindent
The central charge is constant and given by

\begin{equation}
\mathcal{Z}=-\frac{ie^{i\gamma}}{\sqrt{2}}(p^{0}-iq_{0})\equiv 
-\frac{ie^{i\gamma}}{\sqrt{2}}\tilde{q}\, .
\end{equation}

\noindent
The attractor equations do not make sense because $\mathcal{Z}$ is already
moduli-independent. 

The timelike supersymmetric configurations of this theory were first found by
Tod in his pioneering paper Ref.~\cite{Tod:1983pm}, belong to the family
of solutions found by Perj\'es, Israel and Wilson (IWP)
\cite{Perjes:1971gv,kn:IW}; they are completely determined by the choice of
a single complex, harmonic function that we denote by
$\tilde{\mathcal{I}}$. In the framework of general $N=2,d=4$ theories, the
solutions of pure $N=2,d=4$ supergravity are given by just two real harmonic
functions $\mathcal{I}^{0}$ and $\mathcal{I}_{0}$, the components of the real
symplectic vector $\mathcal{I}$. The relation between $\mathcal{I}$ and
$\tilde{\mathcal{I}}$ is

\begin{equation}
\tilde{\mathcal{I}}= \mathcal{I}^{0}-i\mathcal{I}_{0}\, .
\end{equation}

\noindent
Observe that

\begin{equation}
X =  -\frac{ie^{i\gamma}}{\sqrt{2}\tilde{I}}\, ,      
\end{equation}

\noindent
and therefore $\sqrt{2}X$ coincides with the function $V$ of
Ref.~\cite{Tod:1983pm} and is the inverse of the complex harmonic function.

It is convenient to use the complex formulation of this theory. In it,
the symplectic product of two real symplectic vectors $x,y$ can be
written in the form $\langle\, x\mid y\, \rangle =
\Im{\rm m}\, (\tilde{x}^{*}\tilde{y})$ where the tilde indicates
complexification ($\tilde{x}=x^{0}-ix_{0}$ etc.). Further, 
electric-magnetic duality rotations of the symplectic
vectors is equivalent to multiplication by a global phase
$\tilde{x}^{\prime}=e^{i\gamma}\tilde{x}$. We would like to stress that the metric
is invariant under these transformations.

Using Eq.~(\ref{eq:prepotentialpure}) one finds that $\mathcal{R}$,
the real part of $\mathcal{V}/X$ is the symplectic vector

\begin{equation}
\mathcal{R}=
\left(
  \begin{array}{c}
-\mathcal{I}_{0} \\ \mathcal{I}^{0} \\
  \end{array}
\right)\, ,\,\,\, \Rightarrow   
\tilde{\mathcal{R}} = -i\tilde{\mathcal{I}}\, ,
\,\,\, \Rightarrow -g_{rr}= \frac{1}{2|X|^{2}}= 
\langle\, \mathcal{R} \mid \mathcal{I}\, \rangle = 
|\tilde{\mathcal{I}}|^{2}\, .
\end{equation}

\noindent
Finally,

\begin{equation}
\vec{\mathcal{W}} = \tilde{\mathcal{I}}^{*}\vec{\nabla}\tilde{\mathcal{I}}\, .
\end{equation}

It was argued by Hartle and Hawking \cite{Hartle:1972ya} that the only
regular black hole solutions in the IWP family are the static
Majumdar-Papapetrou solutions that describe several charged black holes in
static equilibrium. We are going to see that these are in fact the only
solutions which are everywhere supersymmetric (condition I) and that demanding
positivity of the masses of the components (condition II) is enough to have
regular black holes (condition III plays no r\^{o}le here).

%%%%%%%%%%%%%%%%%%%%%%%%%%%%%%%%%%%%%%%%%%%%%%%%%%%%%%%%%%%%%%%%%%%%%%
%%%%%%%%%%%%%%%%%%%%%%%%%%%%%%%%%%%%%%%%%%%%%%%%%%%%%%%%%%%%%%%%%%%%%%

\subsubsection{Single, static black hole solutions}

The complex harmonic function $\tilde{\mathcal{I}}$ adequate to describe a
static, spherically symmetric, extreme black hole with magnetic and electric
charges $p^{0}$ and $q_{0}$ is

\begin{equation}
\tilde{\mathcal{I}}=\tilde{\mathcal{I}}_{\infty} +\frac{\tilde{q}}{r}\, ,
\hspace{1cm}
\tilde{q}\equiv p^{0}-iq_{0}\, ,
\end{equation}

\noindent
and asymptotic flatness requires $|\tilde{\mathcal{I}}_{\infty}|=1$.
Since $\tilde{\mathcal{I}}_{\infty}$ is just a phase that can be taken 
to be unity by
an electric-magnetic duality rotation. Then,

\begin{equation}
-g_{rr}=|\tilde{\mathcal{I}}|^{2}=1
+\frac{2\Re{\rm e}(\tilde{\mathcal{I}}^{*}_{\infty}\tilde{q} )}{r} 
+\frac{|\tilde{q}|^{2}}{r^{2}}\, . 
\end{equation}

The mass is given by

\begin{equation}
\label{eq:themass}
M= \Re{\rm e} (\tilde{\mathcal{I}}_{\infty}^{*}\tilde{q}) = 
\langle\, \mathcal{R}_{\infty}\mid q\, \rangle\, ,
\end{equation}

\noindent
and the equations of motion and supersymmetry seem to
allow for it to be positive or negative. When $M$ is negative
$|\tilde{I}|^{2}$ will vanish for some finite value of $r$, giving
rise to a naked singularity. In the limit $r\rightarrow 0$, which
makes sense if $M$ is positive, we find that the area of the
2-spheres of constant $t$ and $r$ is finite and equal to

\begin{equation}
A = 4\pi |\tilde{q}|^{2}=8\pi |\mathcal{Z}|^{2}\, .
\end{equation}

\noindent
Observe that, in general,

\begin{equation}
|M| \neq \sqrt{2}|\mathcal{Z}|\, ,   
\end{equation}

\noindent
even though these solutions are usually understood to be supersymmetric.

For this solution
% , the complex vector $\vec{\mathcal{W}}$ is given by
%
% \begin{equation}
% \vec{\mathcal{W}} = (\tilde{\mathcal{I}}_{\infty}^{*}\tilde{q})\, 
% \vec{\nabla}\frac{1}{r}  +|\tilde{q}|^{2}\frac{1}{r}\vec{\nabla}\frac{1}{r}\, ,
% \end{equation}
%
% \noindent
% and
Eq.~(\ref{eq:dW}) is automatically satisfied, while Eq.~(\ref{eq:d*W}) 
takes the form

\begin{equation}
\Im{\rm m}\,  (\vec{\nabla}\cdot \vec{\mathcal{W}})
=  -4\pi \Im{\rm m}\, (\tilde{\mathcal{I}}_{\infty}^{*}\tilde{q})\, 
\delta^{(3)}(\vec{x})=0\, .
\end{equation}

We can, either

\begin{enumerate}
  
\item Adopt the point of view proposed in this paper that the integrability condition
  has to be satisfied everywhere (condition I), whence impose the condition

\begin{equation}
\label{eq:hq}
\Im{\rm  m}\, (\tilde{\mathcal{I}}_{\infty}^{*}\tilde{q})=
\langle\, \mathcal{I}_{\infty} \mid q\, \rangle =  0\, .  
\end{equation}

\noindent
$\tilde{\mathcal{I}}_{\infty}$ is just a phase and this condition
determines it: $\tilde{\mathcal{I}}_{\infty}= \pm
\tilde{q}/|\tilde{q}|\equiv e^{i\beta}$. The complex harmonic function
becomes

\begin{equation}
  \tilde{\mathcal{I}}
  = e^{i\beta}\left(1 \pm \frac{|\tilde{q}|}{r} \right)\, ,
\end{equation}

The overall phase $e^{i\beta}$ is irrelevant for our problem (it can
always be eliminated by an electric-magnetic duality rotation that
does not change the metric), but the relative sign between the two
terms, which is the sign of the mass,

\begin{equation}
M= \pm |\tilde{q}|=\pm |\mathcal{Z}|\, ,  
\end{equation}

\noindent
is important since the minus sign leads to naked singularities. We take the
positive sign as to comply with condition II. We can the integrate the
equation for $\omega$ everywhere. The above condition, however, implies the vanishing 
of the
r.h.s.~of the equation and, therefore, also that of $\omega$. 
Thus, after imposing conditions I and II we obtain a solution which
is static and spherically symmetric and has a regular horizon if
$M>0$; Or

\item We can accept this singularity, ignoring condition I, arguing that,
  after all, the harmonic functions are already singular at that
  point\footnote{We have seen that the solution can, nevertheless, be regular
    at that point, which is the event horizon.} and proceed to integrate the
  equation and obtain $\omega$ which, in spherical coordinates, takes the form

\begin{equation}
% (d\omega)_{\theta\phi}=-\sin{\theta}\langle\, h\mid q\, \rangle\, ,\,\,\,\,
% \Rightarrow\,\,\, 
\omega= 2N \cos{\theta}d\phi\, ,
\end{equation}

\noindent
where $N$ is NUT charge and it is given by

\begin{equation}
\label{eq:NUT}
N=  \Im{\rm   m}\, (\tilde{\mathcal{I}}_{\infty}^{*}\tilde{q})= 
\langle\, \mathcal{I}_{\infty} \mid q\, \rangle\, ,\,\,\,
\Rightarrow |M+iN|=\sqrt{2} |\mathcal{Z}_{\infty}|\, .
\end{equation}

The metric is no longer static, but stationary, and contains either wire
singularities or closed timelike curves plus Taub-NUT asymptotics.

\end{enumerate}
 
It is clear that by imposing conditions I and II, these pathologies are avoided.
Furthermore, in the microscopic models of black holes constructed in the
framework of String Theory there seem to be no configurations that give rise
to macroscopic NUT charge (nor to negative masses). The agreement between
spacetime supersymmetry and the microscopic String Theory models on this point,
together with the elimination of pathologies is encouraging and we will see
that it applies to more cases.

%%%%%%%%%%%%%%%%%%%%%%%%%%%%%%%%%%%%%%%%%%%%%%%%%%%%%%%%%%%%%%%%%%%%%%
%%%%%%%%%%%%%%%%%%%%%%%%%%%%%%%%%%%%%%%%%%%%%%%%%%%%%%%%%%%%%%%%%%%%%%

\subsubsection{Single black hole solutions with a  dipole term}

Let us now consider harmonic functions adequate to describe rotating
supersymmetric black holes. We can add angular momentum to the
previous solution by adding a dipole term to its complex harmonic
function which becomes:

\begin{equation}
\tilde{\mathcal{I}}=\tilde{\mathcal{I}}_{\infty} +\frac{\tilde{q}}{r}
+(\vec{\tilde{m}}\cdot\vec{\nabla})\frac{1}{r}\, ,
\end{equation}

\noindent
where $\vec{m}=(\vec{m}^{0},\vec{m}_{0})$ is a symplectic vector of
dipole magnetic and electric momenta. When they are parallel we can
take them to have only $z$ component and, then, in spherical coordinates

\begin{equation}
\tilde{\mathcal{I}}=\tilde{\mathcal{I}}_{\infty} +\frac{\tilde{q}}{r}
  -\frac{\tilde{m}\cos{\theta}}{r^{2}}\, .
\end{equation}

\noindent
The corresponding $\omega$ (which exists except at the singularities
of $\tilde{\mathcal{I}}$) is

\begin{equation}
\omega= \left[2N \cos{\theta} +2J\frac{\sin^{2}{\theta}}{r^{2}} 
+\Im{\rm m}(\tilde{q}^{*}\tilde{m}) \frac{\sin^{2}{\theta}}{r^{3}}  \right] d\phi\, .
\end{equation}

\noindent
$N$ is the NUT charge and is given again by Eq.~(\ref{eq:NUT}).  The
new features are $J$, the $z$ component of the angular momentum, given
by

\begin{equation}
\label{eq:J}
J =  \Im{\rm   m}\, (\tilde{\mathcal{I}}_{\infty}^{*}\tilde{m})= 
\langle\, \mathcal{I}_{\infty} \mid m, \rangle\, ,
\end{equation}

\noindent
and $\Im{\rm m}(\tilde{q}^{*}\tilde{m})$ which does not have a
conventional name but vanishes when $N=J=0$. 

Let us now analyze the KSIs Eqs.~(\ref{eq:dW}) and (\ref{eq:d*W}) (condition
I). In the general case they take, respectively, the form

\begin{eqnarray}
  2\left[\Im{\rm m}\,(\tilde{q}^{*} \nabla_{m})\vec{\nabla}\frac{1}{r}\right]
\times\vec{\nabla}\frac{1}{r}
-i\left(\nabla_{m^{*}}\vec{\nabla}\frac{1}{r}\right)
\times \left(\nabla_{m}\vec{\nabla}\frac{1}{r}\right)
& = & 0\, ,\\
& & \nonumber \\
\Im{\rm m}\, (\tilde{\mathcal{I}}_{\infty}^{*}\tilde{q})\, \delta^{(3)}(\vec{x})
+
\Im{\rm m}\,(\tilde{\mathcal{I}}_{\infty}^{*}\nabla_{m})\, 
\delta^{(3)}(\vec{x})
+\frac{1}{r}\Im{\rm m}\,(\tilde{q}^{*}\nabla_{m})\, 
\delta^{(3)}(\vec{x})
+ & & \nonumber \\
& & \nonumber \\
+\delta^{(3)}(\vec{x})
\Im{\rm m}\,(\tilde{q}\nabla_{m^{*}})\frac{1}{r}
+\Im{\rm m}\,\left\{\left(\nabla_{m^{*}}\frac{1}{r}\right)
\left(\nabla_{m}\delta^{(3)}(\vec{x})\right)\right\} & = & 0\, ,
\end{eqnarray}

\noindent
and are satisfied if 

\begin{eqnarray}
N & = & \Im{\rm   m}\, (\tilde{\mathcal{I}}_{\infty}^{*}\tilde{q})= 
\langle\, \mathcal{I}_{\infty} \mid q\, \rangle =0\, ,  \\
& & \nonumber \\
\vec{J} & = & \Im{\rm   m}\, (\tilde{\mathcal{I}}_{\infty}^{*}\vec{\tilde{m}})= 
\langle\, \mathcal{I}_{\infty} \mid \vec{m}\, \rangle =0\, ,\\
& & \nonumber \\
& & 
\Im{\rm   m}\, (\tilde{q}^{*}\vec{\tilde{m}})= 
\langle\, q \mid \vec{m}\, \rangle =0\, ,\\
& & \nonumber \\
& & 
\Im{\rm   m}\, (\tilde{m}^{*}_{[\underline{m}}\tilde{m}_{\underline{n}]})= 
\langle\,  m_{[\underline{m}} \mid m_{\underline{n}]}\, \rangle =0\, ,
\end{eqnarray}

\noindent
where we have defined the differential operator $\nabla_{m}\equiv
\vec{\tilde{m}}\cdot\vec{\nabla}$ and where we have taken into account
Eq.~(\ref{eq:J}) to identify the angular momentum.

The first condition is, again, the absence of sources of NUT charge.
The second condition is the absence of sources of angular momentum.
The third and fourth conditions are automatically satisfied in this
theory if the first two are. 

In this case, these conditions are not enough to eliminate all the
singularities introduced by the dipole term since the above conditions
do not cancel terms like
$|\vec{\tilde{m}}\cdot\vec{\nabla}\frac{1}{r}|^{2}$ in the $g_{rr}$
component of the metric and we no longer find a regular 2-sphere in
the $r\rightarrow 0$ limit. However, we are going to argue that,
although technically possible, dipole terms should not be allowed in
$\mathcal{I}$ because their only possible origin is a distribution of
point-like charges and it is the fundamental distribution of
point-like charges that we have to consider in the above equations and
not the field they produce at distances larger than its size. It is in
these conditions that imposing supersymmetry everywhere is equivalent
to cosmic censorship.

Indeed, from the point of view of the electromagnetic fields, the
magnetic dipole momenta, for instance, can have two fundamental
origins: dipole momenta in a distribution of magnetic monopoles or
fundamental dipole momenta that can be seen as stationary electric
currents. In standard electrodynamics the first possibility is
experimentally excluded (see, e.g.~Ref.~\cite{Jackson:1977iu}) but in
$N=2,d=4$ supersymmetric configurations it is the only one allowed
(see Eq.~(\ref{eq:jotamu})).

%%%%%%%%%%%%%%%%%%%%%%%%%%%%%%%%%%%%%%%%%%%%%%%%%%%%%%%%%%%%%%%%%%%%%%
%%%%%%%%%%%%%%%%%%%%%%%%%%%%%%%%%%%%%%%%%%%%%%%%%%%%%%%%%%%%%%%%%%%%%%

\subsubsection{The supersymmetric Kerr-Newman solution}

Therefore we must only consider distributions of static point-like
charges. We will do so in a moment, but there is an interesting
example of rotating black-hole-type solution which must be considered
before: it is given by the complex harmonic function

\begin{equation}
\tilde{\mathcal{I}}=  \tilde{\mathcal{I}}_{\infty} 
+\frac{\tilde{q}}{\tilde{r}}\, ,
\hspace{1cm}
\tilde{r}\equiv \sqrt{x^{2}+y^{2} +(z-i\alpha)^{2}}\, ,
\end{equation}

\noindent
which is known to lead to the (``ultra-extreme'') supersymmetric
Kerr-Newman solution with angular momentum around the $z$ axis;
as is known it has
naked singularities, as all 4-dimensional supersymmetric rotating 
``black-holes'' \cite{Bergshoeff:1996gg}. This is the prototype
of solution for which supersymmetry does not act as a ``cosmic
censor'' as proposed in \cite{Kallosh:1992ii}. Generalizations of this
solution in some other $N=2,d=4$ theories have been constructed in
Ref.~\cite{Behrndt:1997ny}.

The asymptotic expansion of $\tilde{I}$ 

\begin{equation}
\tilde{\mathcal{I}} \sim  \tilde{\mathcal{I}}_{\infty} 
+\frac{\tilde{q}}{r} -\frac{i\alpha\tilde{q}z}{r^{3}}+\cdots\, ,
\end{equation}

\noindent
corresponds to a charge distribution with only two independent
parameters: $\alpha$ and $\tilde{q}$. The magnetic (electric) dipole
momentum is equal to the product of $\alpha$ and the electric
(magnetic) charge and the infinite number of non-vanishing higher
momenta depend also on these few parameters.

According to the point of view advocated here this solution should not be
considered because it corresponds to the far field of a very charge
distribution. As we are going to see, condition I is enough to
exclude it.

Finding the sources of the solution associated to the above complex harmonic
function is very complicated. To start with, $\tilde{\mathcal{I}}$ is singular on the
ring $x^{2}+y^{2}=\alpha^{2}\, ,\,\,\,z=0$ but it is also discontinuous on a
disk bounded by the ring (see e.g.~\cite{Kaiser:2001yn}, whose results we are
going to use here. See also Refs.~\cite{Newman:2002mk,Gsponer:2004gv}.).

% The complex vector $\vec{\mathcal{W}}$ is given by

% \begin{equation}
% \vec{\mathcal{W}} = (\tilde{\mathcal{I}}_{\infty}^{*}\tilde{q})\, 
% \vec{\nabla}\frac{1}{\tilde{r}}
% +|\tilde{q}|^{2}\frac{1}{\tilde{r}^{*}}\vec{\nabla}\frac{1}{\tilde{r}}\, ,
% \end{equation} 

% \noindent
% and the 

Eqs.~(\ref{eq:dW}) and (\ref{eq:d*W}), which express condition I, take,
respectively, the form

\begin{eqnarray}
\Im{\rm m}\, (\tilde{\mathcal{I}}_{\infty}^{*}\tilde{q}) 
\Re{\rm e}\, (\vec{\nabla}\times \vec{C})  
+\Re{\rm e}\, (\tilde{\mathcal{I}}_{\infty}^{*}\tilde{q}) 
\Im{\rm m}\, (\vec{\nabla}\times \vec{C})  
+|\tilde{q}|^{2} \Im{\rm m}\,\left(\frac{1}{\tilde{r}^{*}}
\vec{\nabla}\times \vec{C}\right)
& =  & 0\, ,\\
& & \nonumber \\
\Im{\rm m}\, (\tilde{\mathcal{I}}_{\infty}^{*}\tilde{q}) 
\Re{\rm e}\, (\vec{\nabla}\cdot\vec{C})  
+\Re{\rm e}\, (\tilde{\mathcal{I}}_{\infty}^{*}\tilde{q}) 
\Im{\rm m}\, (\vec{\nabla}\cdot \vec{C})  
+|\tilde{q}|^{2} \Im{\rm m}\,\left(\frac{1}{\tilde{r}^{*}}
\vec{\nabla}\cdot \vec{C}\right)
& = & 0\, , 
\end{eqnarray}

\noindent
where we have defined

\begin{equation}
\vec{C}\equiv \frac{(x,y,z-i\alpha)}{[x^{2}+y^{2} +(z-i\alpha)^{2}]^{3/2}} \, .
\end{equation}

The curl and divergence of $\vec{C}$ have been carefully computed in
Ref.~\cite{Kaiser:2001yn} in a distributional sense, i.e.~as integrals of
their products with test functions. For us it is enough to known that

\begin{equation}
\Re{\rm e}\, (\vec{\nabla}\times \vec{C})= \Im{\rm m}\, (\vec{\nabla}\cdot
\vec{C})=0\, ,
\end{equation}

\noindent
and that $\Im{\rm m}\, (\vec{\nabla}\times \vec{C})$ vanishes for vanishing
$\alpha$. We are left with 

\begin{eqnarray}
\left[\Re{\rm e}\, (\tilde{\mathcal{I}}_{\infty}^{*}\tilde{q}) 
+|\tilde{q}|^{2} \Re{\rm e}\, \frac{1}{\tilde{r}}\right]
\Im{\rm m}\, (\vec{\nabla}\times \vec{C})  
& =  & 0\, ,\\
& & \nonumber \\
\left[\Im{\rm m}\, (\tilde{\mathcal{I}}_{\infty}^{*}\tilde{q}) 
-|\tilde{q}|^{2} \Im{\rm m}\,\frac{1}{\tilde{r}}
\right]
\Re{\rm e}\, (\vec{\nabla}\cdot\vec{C})  
& = & 0\, .
\end{eqnarray}

The only way to satisfy the first condition is to have $\Im{\rm m}\,
(\vec{\nabla}\times \vec{C})=0$, which requires $\alpha=0$ (no sources
of angular momentum). Since $\Re{\rm e}\,
(\vec{\nabla}\cdot\vec{C})\neq 0$ always, the only way to satisfy the
second condition is to have $\Im{\rm m}\,
(\tilde{\mathcal{I}}_{\infty}^{*}\tilde{q})=0$ as before (no sources
of NUT charge) and $\Im{\rm m}\,\frac{1}{\tilde{r}}=0$ which also
requires $\alpha=0$.

Thus, imposing supersymmetry everywhere is equivalent, yet again, to
requiring absence of sources of NUT charge and angular momentum. In
the supersymmetric Kerr-Newman solution all the angular momentum
originates in that source\footnote{We are going to see that there are
  solutions with angular momentum and no elementary sources of angular
  momentum.}  and, thus, that solution and its naked singularities can
be excluded from the class of everywhere supersymmetric solutions of
$N=2,d=4$ supergravity. Again, supersymmetry acts as a cosmic censor
and, most importantly, there is agreement between the macroscopic
description of black holes provided by Supergravity and the
microscopic models provided by String Theory in which there seems to
be no way of having angular momentum without breaking supersymmetry.

Therefore, we must only consider distributions of point-like charges, which
correspond to complex harmonic functions of the form

\begin{equation}
\tilde{\mathcal{I}}=\tilde{\mathcal{I}}_{\infty} 
+\sum_{A}\frac{\tilde{q}_{A}}{|\vec{x}-\vec{x}_{A}|}\, ,
\end{equation}

\noindent
from which dipole (and higher) momenta arise only in asymptotic
expansions:

\begin{equation}
\tilde{\mathcal{I}}\sim \tilde{\mathcal{I}}_{\infty} 
+\frac{\sum_{A}\tilde{q}_{A}}{|\vec{x}|}
+\frac{(\sum_{A}\tilde{q}_{A}\vec{x}_{A})\cdot\vec{x}}{|\vec{x}|^{3}}+\cdots\, ,
\end{equation}

\noindent
and may give rise to non-vanishing angular momentum

\begin{equation}
\vec{J} = \Im{\rm   m}\, (\tilde{\mathcal{I}}_{\infty}^{*}\vec{\tilde{m}})= 
\langle\, \mathcal{I}_{\infty} \mid \vec{m}\, \rangle \, ,
\hspace{1cm}
\vec{m} = -\sum_{A}q_{A}\vec{x}_{A}\, ,
\end{equation}

\noindent
but not to non-vanishing NUT charge.

\begin{equation}
N  = \Im{\rm   m}\, (\tilde{\mathcal{I}}_{\infty}^{*}\tilde{q})= 
\langle\, \mathcal{I}_{\infty} \mid q \, \rangle =0\, ,
\hspace{1cm}
q = \sum_{A}q_{A}\, .
\end{equation}

We are going to look for this kind of solutions in pure $N=2,d=4$ supergravity
next, recovering the (negative) Hartle and Hawking result
\cite{Hartle:1972ya}. We will have to look for them in more general $N=2,d=4$
theories.

%%%%%%%%%%%%%%%%%%%%%%%%%%%%%%%%%%%%%%%%%%%%%%%%%%%%%%%%%%%%%%%%%%%%%%
%%%%%%%%%%%%%%%%%%%%%%%%%%%%%%%%%%%%%%%%%%%%%%%%%%%%%%%%%%%%%%%%%%%%%%

\subsubsection{Solutions with two black holes}
\label{sec-twobhsouren2sugra}

Let us consider, to start with, just two poles

\begin{equation}
\tilde{\mathcal{I}}= \tilde{\mathcal{I}}_{\infty} 
+\frac{\tilde{q}_{1}}{|\vec{x}-\vec{x}_{1}|}
+\frac{\tilde{q}_{2}}{|\vec{x}-\vec{x}_{2}|}\, .
\end{equation}

\noindent
Asymptotic flatness requires $|\tilde{\mathcal{I}}_{\infty}|=1$.  The
condition Eq.~(\ref{eq:dW}) is automatically satisfied and
(\ref{eq:d*W}) takes the form

\begin{equation}
\left[\langle\, \mathcal{I}_{\infty}\mid q_{1}\, \rangle\, 
+\frac{\langle\, q_{2}\mid q_{1}\, \rangle\,}{|\vec{x}_{1}
-\vec{x}_{2}|}\right]\delta^{(3)}(\vec{x}-\vec{x}_{1})
+
\left[\langle\,  \mathcal{I}_{\infty} \mid q_{2}\, \rangle\, 
+\frac{\langle\, q_{1}\mid q_{2}\, \rangle\,}{|\vec{x}_{2}
-\vec{x}_{1}|}\right]\delta^{(3)}(\vec{x}-\vec{x}_{2}) =  0\, ,  
\end{equation}

\noindent
which leads to the two equations

\begin{equation}
\label{eq:twoequations}
  \begin{array}{rcl}
\langle\,  \mathcal{I}_{\infty}\mid q_{1}\, \rangle\, 
+{\displaystyle\frac{\langle\, q_{2}\mid q_{1}\, \rangle\,}{|\vec{x}_{1}
-\vec{x}_{2}|}} & = & 0\, ,\\
& & \\
\langle\,  \mathcal{I}_{\infty} \mid q_{2}\, \rangle\, 
+{\displaystyle\frac{\langle\, q_{1}\mid q_{2}\, \rangle\,}{|\vec{x}_{2}
-\vec{x}_{1}|}} & = & 0\, ,  \\
  \end{array}
\end{equation}

\noindent
each of which expresses the absence of sources of NUT charge at
$\vec{x}_{1}$ and $\vec{x}_{2}$.  The antisymmetry of the symplectic
product implies the consistency condition

\begin{equation}
\label{eq:hq1q2}
\langle\,  \mathcal{I}_{\infty}\mid q_{1}+q_{2}\, \rangle\, =0\, , 
\end{equation}

\noindent
which means that the total charge of the two objects satisfies the
same condition (no global NUT charge) as the charge of just one. 

Expanding asymptotically $\mathcal{I}$ and using the above constraints
we find that this two-body system has a total mass and angular
momentum given by

\begin{eqnarray}
  M & = &  \sum_{A} \langle\,  \mathcal{R}_{\infty}\mid q_{A}\, \rangle \equiv  \sum_{A} M_{A}\, ,\\
  & & \nonumber \\
\label{eq:angularmomentum}
  \vec{J} 
  & = &  
  \langle\, \mathcal{I}_{\infty} \mid \vec{m}\, \rangle 
% = 
% -\langle\, \mathcal{I}_{\infty} \mid q_{A}\, \rangle \vec{x}_{A}
% =
% \langle\, \mathcal{I}_{\infty} \mid q_{1}\, \rangle (\vec{x}_{2}-\vec{x}_{1})
=
\langle\, q_{1} \mid q_{2}\, \rangle \frac{(\vec{x}_{2}-\vec{x}_{1})}{|\vec{x}_{2}-\vec{x}_{1}|}\, .
\end{eqnarray}

Observe that there is total angular momentum even though there are no
sources of angular momentum.

There are two types of solutions to these equations required by condition I:

\begin{enumerate}
\item Each object's charge satisfies the condition for single independent
  objects $\langle\, \mathcal{I}_{\infty} \mid q_{A}\, \rangle =0$ which
  requires $\langle\, q_{2}\mid q_{1}\, \rangle=0$.  In this theory this means
  that the phases of $\tilde{\mathcal{I}}_{\infty},\tilde{q}_{1}$ and
  $\tilde{q}_{2}$ are such that

  \begin{equation}
  \tilde{I} =  e^{i\beta}\left(1 
    +\sum_{A}\frac{s_{A}|\tilde{q}_{A}|}{|\vec{x}-\vec{x}_{A}|} 
\right)\, ,  
  \end{equation}

\noindent
where $s_{A}=\pm 1$. The total mass is given by the formula
Eq.~(\ref{eq:themass})

\begin{equation}
  M= \Re{\rm e} (\tilde{\mathcal{I}}_{\infty}^{*}\sum_{A}\tilde{q}_{A}) = 
  \langle\, \mathcal{R}_{\infty}\mid \sum_{A}q_{A}\, \rangle 
= \sum_{A}s_{A}|\tilde{q}_{A}|\, ,
\end{equation}

\noindent
and the angular momentum vanishes ($\omega$ vanishes).

These are the Majumdar-Papapetrou solutions \cite{Majumdar:1947eu,kn:Pa}. Only
the solutions with all $s_{A}=+1$ are regular, but one could argue that only
those correspond to objects that would have positive masses
$M_{A}=|\tilde{q}_{A}|$ if they were isolated \cite{kn:BrilLin}. This is the
meaning of condition II.

These solutions describe two charged, static black holes in
equilibrium with their event horizons placed at $\vec{x}_{1}$ and
$\vec{x}_{2}$ which are really 2-spheres of finite areas equal to
$4\pi|\tilde{q}_{1}|^{2}$ and $4\pi|\tilde{q}_{2}|^{2}$. They are, as
argued by Hartle and Hawking, and as we are going to see, the only
regular black-hole-type solutions in the whole IWP family
\cite{Hartle:1972ya}

\item $\langle\, \mathcal{I}_{\infty} \mid q_{A}\, \rangle \neq 0$ and
  we have two objects that cannot exist independently in the vacuum
  $\mathcal{I}_{\infty}$ (i.e.~we have a bound state). The distance
  between them is fixed by the condition of absence of sources of NUT
  charge to be

\begin{equation}
|\vec{x}_{2}-\vec{x}_{1}|  =  
{\displaystyle\frac{\langle\, q_{1}\mid q_{2}\, \rangle}
{\langle\, \mathcal{I}_{\infty}\mid q_{1}\, \rangle}}\, .
\end{equation}

\noindent
The sign of the r.h.s.~can always be made positive by flipping the
sign of $\mathcal{I}_{\infty}$, which is irrelevant for the moduli and
for solving Eq.~(\ref{eq:hq1q2}). Thus, this equation always has a
solution. However, when all the above conditions have been satisfied,
the total mass of the solution is negative. The simplest way to see
this is by first making $\tilde{\mathcal{I}}_{\infty}=1$ by a duality rotation
that does not change the metric. After the duality rotation one finds
$\tilde{q}^{\prime}_{A}=M_{A}+iN_{A}$, meaning that they are complex combinations of the
masses and NUT charges of each object. Using $N_{2}=-N_{1}$, the above
condition takes the form

\begin{equation}
N_{1} +\frac{N_{1}M_{2}-N_{2}M_{1}}{|\vec{x}_{2}-\vec{x}_{1}|}=
N_{1}\left(1 +\frac{M_{1}+M_{2}}{|\vec{x}_{2}-\vec{x}_{1}|}\right)=0\, ,  
\end{equation}

\noindent
which has solution only for vanishing NUT charges or for negative total mass
$M_{1}+M_{2}$ which violates condition II and produces naked singularities.
Thus, we cannot simultaneously satisfy conditions I and II for bound states
with 
$\langle\, q_{1}\mid q_{2}\, \rangle \neq 0$.

\end{enumerate}

This result can be generalized to solutions with more poles: let us
consider first the 3-pole harmonic function

\begin{equation}
\tilde{\mathcal{I}}= \tilde{\mathcal{I}}_{\infty} 
+\frac{\tilde{q}_{1}}{|\vec{x}-\vec{x}_{1}|}
+\frac{\tilde{q}_{2}}{|\vec{x}-\vec{x}_{2}|}
+\frac{\tilde{q}_{3}}{|\vec{x}-\vec{x}_{3}|}\, .
\end{equation}

The $\omega$ integrability condition leads to three equations (one to
cancel the NUT charge at each pole) which can be written as a linear
system for the $N_{A}$s:

\begin{equation}
\left(
  \begin{array}{ccc}
   \left(1+\frac{M_{2}}{r_{12}}+\frac{M_{3}}{r_{14}}\right) & -\frac{M_{1}}{r_{12}} &  -\frac{M_{1}}{r_{13}} \\[3mm] 
-\frac{M_{2}}{r_{12}} & \left(1+\frac{M_{1}}{r_{12}}+\frac{M_{3}}{r_{23}}\right) & -\frac{M_{2}}{r_{23}} \\[3mm]
-\frac{M_{3}}{r_{13}} & -\frac{M_{3}}{r_{23}} & \left(1+\frac{M_{1}}{r_{13}}+\frac{M_{2}}{r_{23}}\right) \\
\end{array}
\right)
\left(
  \begin{array}{c}
   N_{1} \\ \\ N_{2} \\ \\ N_{3} \\ 
  \end{array}
\right) =0\, .
\end{equation}

It is easy to see that the determinant of the matrix is $+1$ plus terms linear
and quadratic in the masses, all with positive sign. It will never vanish if
all the masses are positive. This argument can be easily generalized to a
higher number of poles and, therefore we conclude that the only solutions
satisfying conditions I and II are the Majumdar-Papapetrou solutions. This
result should be read in a positive sense: no singular solutions are allowed
by the conditions proposed in the introduction, even if only static solutions
are allowed in this simple theory. To find solutions with angular momentum
satisfying conditions I-III we need to consider theories with scalars.

%%%%%%%%%%%%%%%%%%%%%%%%%%%%%%%%%%%%%%%%%%%%%%%%%%%%%%%%%%%%%%%%%%%%%%
%%%%%%%%%%%%%%%%%%%%%%%%%%%%%%%%%%%%%%%%%%%%%%%%%%%%%%%%%%%%%%%%%%%%%%

\subsection{General $N=2,d=4$ supergravity}
\label{sec-generaln2d4}

The setup of our problem in general $N=2,d=4$ theories is similar to pure
supergravity case. Let us first consider spherically-symmetric, static, single
black-hole-type solutions with magnetic and electric charges $p^{\Lambda}$ and
$q_{\Lambda}$. They are determined by a symplectic vector of $2\bar{n}$ real
harmonic functions

\begin{equation}
\mathcal{I}=
\left(
  \begin{array}{c}
\mathcal{I}^{\Lambda} \\ \mathcal{I}_{\Lambda} \\
  \end{array}
\right)
=
\mathcal{I}_{\infty} +\frac{q}{r}\, ,\
\hspace{1cm}
q\equiv
\left(
  \begin{array}{c}
p^{\Lambda} \\ q_{\Lambda} \\
  \end{array}
\right)\, ,
\hspace{1cm}
\mathcal{I}_{\infty}
\equiv
\left(
  \begin{array}{c}
\mathcal{I}_{\infty}^{\Lambda} \\ \mathcal{I}_{\Lambda\, \infty} \\
  \end{array}
\right)\, .
\end{equation}

We assume that the stabilization equations have been solved and
$\mathcal{R}(\mathcal{I})$ has been found in order to be able to construct the
fields of the solutions. 

The $n$ complex scalars are constructed using the general formula
Eq.~(\ref{eq:scalars}). The moduli (the values of the $n$ complex scalars
$Z^{i}$ at infinity, $Z^{i}_{\infty}$) are complicated functions
$Z^{i}_{\infty}(\mathcal{I}_{\infty})$ of these $2n+2$ real constant
components of $\mathcal{I}_{\infty}$.  One of the components of
$\mathcal{I}_{\infty}$ can be determined as a function of the remaining $2n+1$
by imposing asymptotic flatness of the metric, that is, $\langle\,
\mathcal{R}_{\infty} \mid \mathcal{I}_{\infty}\, \rangle=1$, and another one
can be determined by imposing condition I, since Eq.~(\ref{eq:iddi}) implies

\begin{equation}
N=\langle\, \mathcal{I}_{\infty}\mid q\, \rangle=0\, .
\end{equation}

\noindent
It should always be possible to give the $2n$ real moduli any admissible value
within their definition domain with the remaining $2n$ unconstrained real
components of $\mathcal{I}_{\infty}$. This is difficult to prove explicitly
due to the complicated and theory-dependent relations between
$\mathcal{I}_{\infty}$ and the moduli $Z^{i}_{\infty}$, but it is safe to
assume that in general it is possible.

Let us turn to condition II. The positivity of the masses, which is given by
the general expression Eq.~(\ref{eq:linearBPSmass}) has to be imposed by hand
and, although this can always be done, it is a non-trivial constraint on the
charges and moduli. The positivity of the masses can be also understood as
part of a stronger requirement that the scalar fields take values only within
their definition domain for all values of $r$.  Actually, this requirement
should suffice to ensure the finiteness of $-g_{rr}$ for $r\neq 0$.

The finiteness of $-g_{rr}$ for $r\neq 0$ is not enough to have a black hole
and condition III has to be imposed to find a finite horizon area at $r=0$.

If we want to describe more than one black hole we have to use harmonic
functions with two point-like singularities:

\begin{equation}
\mathcal{I}=\mathcal{I}_{\infty} +\frac{q_{1}}{|\vec{x}-\vec{x}_{1}|}
+\frac{q_{2}}{|\vec{x}-\vec{x}_{2}|}\, .
\end{equation}

\noindent
Again, one of the components of $\mathcal{I}_{\infty}$ is determined by
imposing asymptotic flatness. Condition I now leads to the two equations
Eqs.~(\ref{eq:twoequations}) which should determine another component of
$\mathcal{I}_{\infty}$ and the parameter $|\vec{x}_{1}-\vec{x}_{2}|$ if
$\langle q_{2} \mid q_{1}\rangle \neq 0$. The question now is whether these
solutions can be obtained while maintaining the positivity of the masses
(condition II) 

\begin{equation}
M_{i} \equiv \langle\, \mathcal{R}_{\infty} \mid q_{i}\, \rangle >0\, ,  
\end{equation}

\noindent
and solving the attractor equations for each of the singularities of the
harmonic functions. We have no general answer to these questions and, what we
are going to do is to study how the three conditions can actually be imposed
in a particularly simple example and suffice to ensure regularity of the
solutions.

%%%%%%%%%%%%%%%%%%%%%%%%%%%%%%%%%%%%%%%%%%%%%%%%%%%%%%%%%%%%%%%%%%%%%%
%%%%%%%%%%%%%%%%%%%%%%%%%%%%%%%%%%%%%%%%%%%%%%%%%%%%%%%%%%%%%%%%%%%%%%
%%%%%%%%%%%%%%%%%%%%%%%%%%%%%%%%%%%%%%%%%%%%%%%%%%%%%%%%%%%%%%%%%%%%%%
%%%%%%%%%%%%%%%%%%%%%%%%%%%%%%%%%%%%%%%%%%%%%%%%%%%%%%%%%%%%%%%%%%%%%%
%%%%%%%%%%%%%%%%%%%%%%%%%%%%%%%%%%%%%%%%%%%%%%%%%%%%%%%%%%%%%%%%%%%%%%
%%%%%%%%%%%%%%%%%%%%%%%%%%%%%%%%%%%%%%%%%%%%%%%%%%%%%%%%%%%%%%%%%%%%%%

\subsubsection{A toy model with a complex scalar field}
\label{sec-toy}

We are going to consider the $\bar{n}=2$ theory with prepotential

\begin{equation}
\mathcal{F}=-i\mathcal{X}^{0}\mathcal{X}^{1}\, .  
\end{equation}

\noindent
This theory has only one complex scalar 

\begin{equation}
\tau\equiv i\mathcal{X}^{1}/\mathcal{X}^{0}\, , 
\end{equation}

\noindent
in terms of which the period matrix is given by 

\begin{equation}
(\mathcal{N}_{\Lambda\Sigma})=
\left(
  \begin{array}{cc}
-\tau & 0 \\
0 & 1/\tau \\
  \end{array}
\right)  
\end{equation}

\noindent
and, in the $\mathcal{X}^{0}=i/2$ gauge, the K\"ahler potential 
and metric are

\begin{equation}
\mathcal{K}=-\ln{\Im{\rm m}\tau}\, ,  
\hspace{1cm}
\mathcal{G}_{\tau\tau^{*}} = (2\Im{\rm m}\tau)^{-2}\, .
\end{equation}

\noindent
The reality of the K\"ahler potential requires the positivity of $\Im{\rm
  m}\tau$. Therefore, $\tau$ parametrizes the coset $SL(2,\mathbb{R})/SO(2)$
and can be identified with the \textit{axidilaton} and this theory is a
truncation of the $SO(4)$ formulation of $N=4,d=4$ supergravity.

The symplectic section $\mathcal{V}$ is 

\begin{equation}
\mathcal{V}=\frac{1}{2(\Im {\rm m}\tau)^{1/2}}
\left( 
  \begin{array}{c}
i \\ \tau \\ -i\tau \\ 1 \\
  \end{array}
\right)\, ,  
\end{equation}

\noindent
and the central charge is

\begin{equation}
\mathcal{Z}(\tau,\tau^{*},q)= \langle \mathcal{V} \mid q \rangle =
\frac{1}{2(\Im {\rm m}\tau)^{1/2}}[(p^{1}-iq_{0}) -(q_{1}+ip^{0})\tau]\, .  
\end{equation}

The attractor equation is

\begin{equation}
\left. \frac{d}{d\tau} \frac{1}{\Im {\rm m}\tau}[(p^{1}-iq_{0})
-(q_{1}+ip^{0})\tau]\right|_{\tau=\tau_{\rm fix}}=0\, , 
\end{equation}

\noindent
and has the general solution 

\begin{equation}
\tau_{\rm fix} = \frac{p^{1}+iq_{0}}{q_{1}-ip^{0}}\, ,  
\end{equation}

\noindent
which is admissible (belongs to the definition domain of $\tau$) if

\begin{equation}
\Im{\rm m}\tau_{\rm fix} = p^{0}p^{1}+q_{0}q_{1} > 0\, . 
\end{equation}

\noindent
The central charge at the fixed point of the scalar takes the value

\begin{equation}
\mathcal{Z}_{\rm fix} = -i \frac{q_{1}+ip^{0}}{|q_{1}+ip^{0}|} 
\sqrt{p^{0}p^{1}+q_{0}q_{1}}\, , 
\end{equation}

\noindent
and it is always finite for $\tau_{\rm fix} \neq 0$.

%%%%%%%%%%%%%%%%%%%%%%%%%%%%%%%%%%%%%%%%%%%%%%%%%%%%%%%%%%%%%%%%%%%%%%
%%%%%%%%%%%%%%%%%%%%%%%%%%%%%%%%%%%%%%%%%%%%%%%%%%%%%%%%%%%%%%%%%%%%%%
\vspace{.3cm}
\noindent
\underline{\textit{Solutions with a single black hole}}
\vspace{.3cm}

Let us now consider solutions with

\begin{equation}
\mathcal{I}=\mathcal{I}_{\infty} +\frac{q}{r}\, .
\end{equation}

\noindent
In this theory the stabilization equations can be easily solved and
they lead to 

\begin{equation}
\mathcal{R}= 
\left(
  \begin{array}{cc}
0 & -\sigma^{1} \\
\sigma^{1} & 0 \\
  \end{array}
\right)\mathcal{I}\, ,
\,\,\,\,
\Rightarrow
\,\,\,\,
-g_{rr}=\langle\, \mathcal{R} \mid \mathcal{I}\, \rangle =
2(\mathcal{I}^{0}\mathcal{I}^{1}+\mathcal{I}_{0}\mathcal{I}_{1})\, ,
\end{equation}

\noindent
which shows that the area of the horizon (if any) is related to
$|\mathcal{Z}_{\rm fix}|^{2}$ above according to the general formula
Eq.~(\ref{eq:AZfix}).

We also have

\begin{equation}
\tau = i\frac{\mathcal{L}^{1}/X}{\mathcal{L}^{0}/X}= 
\frac{\mathcal{I}^{1}+i\mathcal{I}_{0}}{\mathcal{I}_{1}-i\mathcal{I}^{0}}\, ,
\end{equation}

\noindent
which implies that the 4 harmonic functions are not entirely independent but
have to satisfy

\begin{equation}
\label{eq:esacondicion}
\Im{\rm m}\tau  = \mathcal{I}^{0} \mathcal{I}^{1}
+\mathcal{I}_{0} \mathcal{I}_{1} > 0\, , 
\end{equation}

\noindent
which ensures that, if there are no pathologies that make a black-hole
interpretation of the solution impossible, the attractor equations will always
have solutions and $\mathcal{Z}_{\rm fix}\neq 0$. Thus, we will not have to
worry about condition III but only about the positive definiteness of $\Im{\rm
  m}\tau$. 

The only possible pathologies (negative mass and presence of NUT charge) are
clearly avoided by imposing conditions I and II, which is always possible and
presents no difficulties.

%%%%%%%%%%%%%%%%%%%%%%%%%%%%%%%%%%%%%%%%%%%%%%%%%%%%%%%%%%%%%%%%%%%%%%
%%%%%%%%%%%%%%%%%%%%%%%%%%%%%%%%%%%%%%%%%%%%%%%%%%%%%%%%%%%%%%%%%%%%%%
\vspace{.3cm}
\noindent
\underline{\textit{Solutions with two black holes}}
\vspace{.3cm}

Let us now consider solutions of the form

\begin{equation}
\mathcal{I}=\mathcal{I}_{\infty} +\frac{q_{1}}{r_{1}}
+\frac{q_{2}}{r_{2}}\, ,
\hspace{1cm}
r_{i}\equiv |\vec{x}-\vec{x}_{i}|\, .
\end{equation}

Our goal is to find a configuration (i.e.~a set of asymptotic values
$\mathcal{I}_{\infty}$ and charges $q_{1,2}$) that satisfy conditions I-III.
The previous discussions indicate how this has to be done and which formulas
need to be applied. There is no systematic procedure to find such a
configuration but it is not too difficult to find one:

\begin{equation}
  \begin{array}{rcl}
\mathcal{I}^{0} & = & \,\,\,\,\,\,
{\displaystyle
\frac{1}{\sqrt{2}} 
+\frac{q}{r_{1}}+\frac{q}{r_{2}}
}\, ,\\
& & \\
\mathcal{I}^{1} & = &  \,\,\,\,\,\, 
{\displaystyle
\frac{1}{\sqrt{2}} 
+\frac{8q}{r_{1}}+\frac{8q}{r_{2}}
}\, ,\\
& & \\
\mathcal{I}_{0} & = &  \hspace{2.2cm} 
{\displaystyle-\frac{4q}{r_{2}}
}\, ,\\
& & \\
\mathcal{I}_{1} & = & 
{\displaystyle
-\frac{1}{4\sqrt{2}} 
-\frac{q}{r_{1}}+\frac{q}{r_{2}}
}\, ,\\
  \end{array}
\end{equation}

\noindent
where $q>0$ in order to guarantee Eq.~(\ref{eq:esacondicion}).  The metric
component

\begin{equation}
-g_{rr}= 1 +\frac{9\sqrt{2}q}{r_{1}}  
+\frac{10\sqrt{2}q}{r_{2}}
+\frac{16q^{2}}{r_{1}^{2}}  
+\frac{8q^{2}}{r_{2}^{2}}
+\frac{40q^{2}}{r_{1}r_{2}}\, ,
\end{equation}

\noindent
is finite everywhere outside $r_{1,2}=0$, and therefore, so is $\Im{\rm
  m}\tau$. In particular the ``mass'' of each of the two objects is positive

\begin{equation}
M_{1}=9q/\sqrt{2}\, ,
\hspace{1cm}
M_{2}=5\sqrt{2}q\, ,  
\hspace{1cm}
M=M_{1}+M_{2}=19q/\sqrt{2}\, ,
\end{equation}

\noindent
and in the $r_{1,2}\rightarrow 0$ limits we find spheres of finite areas

\begin{equation}
\frac{A_{1}}{4\pi} = 16q^{2}= 2|\mathcal{Z}_{\rm fix,1}|^{2}\, ,
\hspace{1cm}
\frac{A_{2}}{4\pi} = 8q^{2} = 2|\mathcal{Z}_{\rm fix,2}|^{2}\, .  
\end{equation}

\noindent
The total horizon area is

\begin{equation}
\frac{A}{4\pi}=\frac{A_{1}}{4\pi}+\frac{A_{2}}{4\pi} = 24q^{2} < 
 2|\mathcal{Z}_{\rm fix, tot}|^{2}=64q^{2}\, ,
\end{equation}

\noindent
which is the area of the horizon of a single black hole having the sum of the
charges of the two black holes.

For this configuration

\begin{equation}
\langle\, \mathcal{I}_{\infty} \mid q_{1}\, \rangle = 
-\langle\, \mathcal{I}_{\infty} \mid q_{2}\, \rangle =-q/\sqrt{2}\, , 
\hspace{1cm}
\langle\, q_{2} \mid q_{1}\, \rangle= 12q^{2}\, ,
\end{equation}

\noindent
so, choosing 

\begin{equation}
r_{12}=|\vec{x}_{2}-\vec{x}_{1}|=12\sqrt{2}q\, ,   
\end{equation}

\noindent
we satisfy condition I (no NUT charges). The system has nevertheless
angular momentum given by the general formula Eq.~(\ref{eq:angularmomentum}):

\begin{equation}
|J|= |\langle\, q_{2} \mid q_{1}\, \rangle| = 12q^{2}\, . 
\end{equation}

%%%%%%%%%%%%%%%%%%%%%%%%%%%%%%%%%%%%%%%%%%%%%%%%%%%%%%%%%%%%%%%%%%%%%%
%%%%%%%%%%%%%%%%%%%%%%%%%%%%%%%%%%%%%%%%%%%%%%%%%%%%%%%%%%%%%%%%%%%%%%
%%%%%%%%%%%%%%%%%%%%%%%%%%%%%%%%%%%%%%%%%%%%%%%%%%%%%%%%%%%%%%%%%%%%%%
%%%%%%%%%%%%%%%%%%%%%%%%%%%%%%%%%%%%%%%%%%%%%%%%%%%%%%%%%%%%%%%%%%%%%%
%%%%%%%%%%%%%%%%%%%%%%%%%%%%%%%%%%%%%%%%%%%%%%%%%%%%%%%%%%%%%%%%%%%%%%
%%%%%%%%%%%%%%%%%%%%%%%%%%%%%%%%%%%%%%%%%%%%%%%%%%%%%%%%%%%%%%%%%%%%%%

\section{Conclusions}
\label{sec-conclusion}

We have formulated three conditions that supersymmetric black-hole-type
solutions have to satisfy in order to be supersymmetric everywhere, including
at the sources. We have shown how these conditions constrain the possible
sources by, basically, excluding those with NUT charge, angular momentum, negative energy
and scalar hair, which seemingly cannot be modeled in String Theory.
We arrived
at a picture in which if an observer far away from one of the globally
supersymmetric configurations we have considered, detects angular momentum
and non-trivial scalar fields he/she will only find static electromagnetic
sources in equilibrium when approaching the system.

These conditions and this picture should be improved by considering quantum
corrections. Another interesting course of action would be to consider regularity
of black-hole solutions in $N>2$ theories, e.g.~\cite{Cvetic:1995uj}, and 
investigate the r\^{o}le played by the attractor \cite{Ferrara:1996um}.
 
It is also clear that the situation in $d=5$ is completely different
as there are regular rotating supersymmetric black holes for which
microscopic String Theory models are known \cite{Breckenridge:1996is}. Work on
these issues is already in progress \cite{kn:BMO}.

%%%%%%%%%%%%%%%%%%%%%%%%%%%%%%%%%%%%%%%%%%%%%%%%%%%%%%%%%%%%%%%%%%%%%%
%%%%%%%%%%%%%%%%%%%%%%%%%%%%%%%%%%%%%%%%%%%%%%%%%%%%%%%%%%%%%%%%%%%%%%
%%%%%%%%%%%%%%%%%%%%%%%%%%%%%%%%%%%%%%%%%%%%%%%%%%%%%%%%%%%%%%%%%%%%%%
%%%%%%%%%%%%%%%%%%%%%%%%%%%%%%%%%%%%%%%%%%%%%%%%%%%%%%%%%%%%%%%%%%%%%%
%%%%%%%%%%%%%%%%%%%%%%%%%%%%%%%%%%%%%%%%%%%%%%%%%%%%%%%%%%%%%%%%%%%%%%

\section*{Acknowledgments}

T.O.~would like to thank Renata Kallosh for pointing the authors towards
references \cite{Denef:2000nb} and \cite{Bates:2003vx} and for useful
conversations, Roberto Emparan for his explanations concerning rotating
black-hole configurations and many other useful comments and, finally,
M.M.~Fern\'andez for her long standing support.

This work has been supported in part by the Spanish Ministry of Science and
Education grant BFM2003-01090 and the Comunidad de Madrid grant HEPHACOS
P-ESP-00346.

%%%%%%%%%%%%%%%%%%%%%%%%%%%%%%%%%%%%%%%%%%%%%%%%%%%%%%%%%%%%%%%%%%%%%%
%%%%%%%%%%%%%%%%%%%%%%%%%%%%%%%%%%%%%%%%%%%%%%%%%%%%%%%%%%%%%%%%%%%%%%
%%%%%%%%%%%%%%%%%%%%%%%%%%%%%%%%%%%%%%%%%%%%%%%%%%%%%%%%%%%%%%%%%%%%%%
%%%%%%%%%%%%%%%%%%%%%%%%%%%%%%%%%%%%%%%%%%%%%%%%%%%%%%%%%%%%%%%%%%%%%%
%%%%%%%%%%%%%%%%%%%%%%%%%%%%%%%%%%%%%%%%%%%%%%%%%%%%%%%%%%%%%%%%%%%%%%
%%%%%%%%%%%%%%%%%%%%%%%%%%%%%%%%%%%%%%%%%%%%%%%%%%%%%%%%%%%%%%%%%%%%%%
\appendix
%%%%%%%%%%%%%%%%%%%%%%%%%%%%%%%%%%%%%%%%%%%%%%%%%%%%%%%%%%%%%%%%%%%%%%
%%%%%%%%%%%%%%%%%%%%%%%%%%%%%%%%%%%%%%%%%%%%%%%%%%%%%%%%%%%%%%%%%%%%%%
%%%%%%%%%%%%%%%%%%%%%%%%%%%%%%%%%%%%%%%%%%%%%%%%%%%%%%%%%%%%%%%%%%%%%%
%%%%%%%%%%%%%%%%%%%%%%%%%%%%%%%%%%%%%%%%%%%%%%%%%%%%%%%%%%%%%%%%%%%%%%
%%%%%%%%%%%%%%%%%%%%%%%%%%%%%%%%%%%%%%%%%%%%%%%%%%%%%%%%%%%%%%%%%%%%%%

\section{Proofs of some identities}
\label{sec-proofs}

Let us consider the generalized stabilization equations derived from
Eq.~(\ref{eq:MdF}). Differentiating the imaginary part of that equation
(i.e.~Eq.(\ref{eq:IRHH}), we get

\begin{equation}
d\mathcal{I}_{\Lambda}= d\Im{\rm m}\mathcal{F}_{\Lambda} =
{\textstyle\frac{1}{2i}}(d\mathcal{X}^{\Lambda}\mathcal{F}_{\Sigma\Lambda}
-d\mathcal{X}^{*\, \Lambda}\mathcal{F}^{*}_{\Sigma\Lambda})=
d\mathcal{R}^{\Sigma}\Im{\rm m}\mathcal{F}_{\Sigma\Lambda}
+d\mathcal{I}^{\Sigma}\Re{\rm e}\mathcal{F}_{\Sigma\Lambda}\, ,
\end{equation}

\noindent
where we have used
$\mathcal{X}^{\Lambda}=\mathcal{R}^{\Lambda}+i\mathcal{I}^{\Lambda}$.  Using
the invertibility of the imaginary part of $\mathcal{F}_{\Sigma\Lambda}$ we
get

\begin{equation}
d\mathcal{R}^{\Sigma} = 
\Im{\rm m}\mathcal{F}^{\Sigma\Lambda}d\mathcal{I}_{\Lambda}
-\Im{\rm m}\mathcal{F}^{\Sigma\Omega}
\Re{\rm   e}\mathcal{F}_{\Omega\Lambda}d\mathcal{I}^{\Lambda}\, .
\end{equation}

On the other hand, differentiating the real part of Eq.~(\ref{eq:MdF})

\begin{equation}
d\mathcal{R}_{\Lambda}= d\Re{\rm e}\mathcal{F}_{\Lambda}= 
{\textstyle\frac{1}{2}}(d\mathcal{X}^{\Lambda}\mathcal{F}_{\Sigma\Lambda}
+d\mathcal{X}^{*\, \Lambda}\mathcal{F}^{*}_{\Sigma\Lambda})=  
d\mathcal{R}^{\Sigma}\Re{\rm e}\mathcal{F}_{\Sigma\Lambda}
-d\mathcal{I}^{\Sigma}\Im{\rm m}\mathcal{F}_{\Sigma\Lambda}\, ,
\end{equation}

\noindent
and, substituting our previous result for $d\mathcal{R}^{\Lambda}$

\begin{equation}
d\mathcal{R}_{\Sigma}= \Re{\rm e}\mathcal{F}_{\Sigma\Omega}
\Im{\rm m}\mathcal{F}^{\Omega\Lambda}dH_{\Lambda}
-(\Im{\rm m}\mathcal{F}_{\Sigma\Lambda} 
+\Re{\rm e}\mathcal{F}_{\Sigma\Omega}\Im{\rm m}\mathcal{F}^{\Omega\Delta}
\Re{\rm   e}\mathcal{F}_{\Delta\Lambda})d\mathcal{I}^{\Lambda}\, .
\end{equation}

We can write all these results in the form

\begin{eqnarray}
d\mathcal{R} & = & 
\left(
  \begin{array}{cc}
-\Im{\rm m}\mathcal{F}^{-1}\Re{\rm e}\mathcal{F} & \Im{\rm m}\mathcal{F}^{-1}
 \\
& \\
-(\Im{\rm m}\mathcal{F} 
+\Re{\rm e}\mathcal{F}\Im{\rm m}\mathcal{F}^{-1}\Re{\rm e}\mathcal{F}) &
\Re{\rm e}\mathcal{F}\Im{\rm m}\mathcal{F}^{-1}
  \end{array}
\right)
d\mathcal{I}\, , \\
& & \nonumber \\
& & \nonumber \\
d\mathcal{I} & = & 
\left(
  \begin{array}{cc}
\Im{\rm m}\mathcal{F}^{-1}\Re{\rm e}\mathcal{F} & -\Im{\rm m}\mathcal{F}^{-1}
 \\
& \\
\Im{\rm m}\mathcal{F} 
+\Re{\rm e}\mathcal{F}\Im{\rm m}\mathcal{F}^{-1}\Re{\rm e}\mathcal{F} &
-\Re{\rm e}\mathcal{F}\Im{\rm m}\mathcal{F}^{-1}
  \end{array}
\right)
d\mathcal{R}\, , 
\end{eqnarray}

\noindent
from which we can read identities such as

\begin{equation}
  \begin{array}{rclclrclcl}
{\displaystyle
\frac{\partial \mathcal{R}^{\Sigma}}{\partial \mathcal{I}_{\Lambda}}
}  
& = & 
{\displaystyle
\frac{\partial \mathcal{R}^{\Lambda}}{\partial \mathcal{I}_{\Sigma}} 
}
& = & 
{\displaystyle
\frac{\partial \mathcal{I}^{\Lambda}}{\partial \mathcal{R}_{\Sigma}}\, ,
}
\hspace{1.5cm} &  
{\displaystyle
\frac{\partial \mathcal{R}_{\Sigma}}{\partial \mathcal{I}^{\Lambda}} 
}
& = &
{\displaystyle
\frac{\partial \mathcal{R}_{\Lambda}}{\partial \mathcal{I}^{\Sigma}} 
}
& = &
{\displaystyle
-\frac{\partial \mathcal{I}_{\Lambda}}{\partial \mathcal{R}^{\Sigma}}
\, , 
}\\
& & & & & & & \\
{\displaystyle
\frac{\partial \mathcal{R}^{\Sigma}}{\partial \mathcal{I}^{\Lambda}}
}
 & = &
{\displaystyle
-\frac{\partial \mathcal{R}_{\Lambda}}{\partial \mathcal{I}_{\Sigma}}
}
& = & 
{\displaystyle
\frac{\partial \mathcal{I}_{\Lambda}}{\partial \mathcal{R}_{\Sigma}}
\, ,
}
&   
{\displaystyle
\frac{\partial \mathcal{R}_{\Sigma}}{\partial \mathcal{I}_{\Lambda}}
}
 & = & 
{\displaystyle
-\frac{\partial \mathcal{R}^{\Lambda}}{\partial \mathcal{I}^{\Sigma}} 
}
& = & 
{\displaystyle
\frac{\partial \mathcal{I}^{\Lambda}}{\partial \mathcal{R}^{\Sigma}}\, . 
}
\\
\end{array}
\end{equation}

We can now prove Eq.~(\ref{eq:dRI=RdI}): taking the derivative of
$\mathcal{R}$ as a function of $\mathcal{I}$ we have

\begin{equation}
  \begin{array}{rcl}
 \langle\, \nabla_{\mu}\mathcal{R} \mid 
\mathcal{I} \, \rangle & = & 
{\displaystyle
\langle\,
\frac{\partial \mathcal{R}}{\partial \mathcal{I}^{\Lambda}}\nabla_{\mu} \mathcal{I}^{\Lambda}  
+\frac{\partial \mathcal{R}}{\partial \mathcal{I}_{\Lambda}}\nabla_{\mu} \mathcal{I}_{\Lambda}  
\mid \mathcal{I} \, \rangle 
}
\\
& & \\
& =  & 
{\displaystyle
\nabla_{\mu}\mathcal{I}^{\Lambda}
\left(
\mathcal{I}^{\Sigma}
\frac{\partial \mathcal{R}_{\Sigma}}{\partial \mathcal{I}^{\Lambda}} 
-
\mathcal{I}_{\Sigma}
\frac{\partial \mathcal{R}^{\Sigma}}{\partial \mathcal{I}^{\Lambda}} 
\right)
+
\nabla_{\mu}\mathcal{I}_{\Lambda}
\left(
\mathcal{I}^{\Sigma}
\frac{\partial \mathcal{R}_{\Sigma}}{\partial \mathcal{I}_{\Lambda}} 
-
\mathcal{I}_{\Sigma}
\frac{\partial \mathcal{R}^{\Sigma}}{\partial \mathcal{I}_{\Lambda}} 
\right)\, ,
}
\\
\end{array}
\end{equation}

\noindent
and using now the above relations between partial derivatives

\begin{equation}
 \langle\, \nabla_{\mu}\mathcal{R} \mid 
\mathcal{I} \, \rangle  =  
\nabla_{\mu}\mathcal{I}^{\Lambda}
\left(
\mathcal{I}^{\Sigma}
\frac{\partial \mathcal{R}_{\Lambda}}{\partial \mathcal{I}^{\Sigma}} 
+
\mathcal{I}_{\Sigma}
\frac{\partial \mathcal{R}_{\Lambda}}{\partial \mathcal{I}_{\Sigma}} 
\right)
-
\nabla_{\mu}\mathcal{I}_{\Lambda}
\left(
\mathcal{I}^{\Sigma}
\frac{\partial \mathcal{R}^{\Lambda}}{\partial \mathcal{I}^{\Sigma}} 
+
\mathcal{I}_{\Sigma}
\frac{\partial \mathcal{R}^{\Lambda}}{\partial \mathcal{I}_{\Sigma}} 
\right)\, .
\end{equation}

\noindent
Given that the real section $\mathcal{R}$ is homogeneous of first order in the
$\mathcal{I}$'s

\begin{equation}
\mathcal{I}^{\Sigma}
\frac{\partial \mathcal{R}_{\Lambda}}{\partial \mathcal{I}^{\Sigma}} 
+
\mathcal{I}_{\Sigma}
\frac{\partial \mathcal{R}_{\Lambda}}{\partial \mathcal{I}_{\Sigma}} 
=  \mathcal{R}_{\Lambda}\, ,
\hspace{2cm}
\mathcal{I}^{\Sigma}
\frac{\partial \mathcal{R}^{\Lambda}}{\partial \mathcal{I}^{\Sigma}} 
+
\mathcal{I}_{\Sigma}
\frac{\partial \mathcal{R}^{\Lambda}}{\partial \mathcal{I}_{\Sigma}} 
=  \mathcal{R}^{\Lambda}\, ,
\end{equation}

\noindent
which proves the identity.

Similarly, expanding the r.h.s.~of Eq.~(\ref{eq:RdR=IdI}) we get

\begin{equation}
 \langle\, \mathcal{R} \mid \nabla_{\mu}\mathcal{R} \, \rangle  =
\left(
\frac{\partial\mathcal{R}^{\Lambda}}{\partial\mathcal{I}^{\Sigma}}
R_{\Lambda} 
-
\frac{\partial\mathcal{R}_{\Lambda}}{\partial\mathcal{I}^{\Sigma}}
R^{\Lambda} 
\right)d\mathcal{I}^{\Sigma}
+
\left(
\frac{\partial\mathcal{R}^{\Lambda}}{\partial\mathcal{I}_{\Sigma}}
R_{\Lambda} 
-
\frac{\partial\mathcal{R}_{\Lambda}}{\partial\mathcal{I}_{\Sigma}}
R^{\Lambda} 
\right)d\mathcal{I}_{\Sigma}\, ,
\end{equation}

\noindent
and using the identities between partial derivatives and the fact that the
real section $\mathcal{I}$ is homogeneous of first order in $\mathcal{R}$, we
arrive at the result we wanted.

%%%%%%%%%%%%%%%%%%%%%%%%%%%%%%%%%%%%%%%%%%%%%%%%%%%%%%%%%%%%%%%%%%%%%%
%%%%%%%%%%%%%%%%%%%%%%%%%%%%%%%%%%%%%%%%%%%%%%%%%%%%%%%%%%%%%%%%%%%%%%
%%%%%%%%%%%%%%%%%%%%%%%%%%%%%%%%%%%%%%%%%%%%%%%%%%%%%%%%%%%%%%%%%%%%%%
%%%%%%%%%%%%%%%%%%%%%%%%%%%%%%%%%%%%%%%%%%%%%%%%%%%%%%%%%%%%%%%%%%%%%%
%%%%%%%%%%%%%%%%%%%%%%%%%%%%%%%%%%%%%%%%%%%%%%%%%%%%%%%%%%%%%%%%%%%%%%
%%%%%%%%%%%%%%%%%%%%%%%%%%%%%%%%%%%%%%%%%%%%%%%%%%%%%%%%%%%%%%%%%%%%%%

\end{document}